\documentclass[12pt]{article}
\usepackage[colorlinks=true,linkcolor=black,anchorcolor=black,citecolor=black,filecolor=black,menucolor=black,urlcolor=black]{hyperref}
\usepackage{amssymb,amsmath,amsfonts}
\usepackage{color}

\newcommand{\com}[1]{{\color{blue} #1 \color{black}}}
\usepackage{latexsym}
\usepackage{graphicx}
\usepackage{psfrag}
\usepackage{mathrsfs}
\usepackage[Symbol]{upgreek}
\usepackage{bbm}
\usepackage{dsfont}
\usepackage{textcomp}
\usepackage{wasysym}
\usepackage{graphicx}
\usepackage{caption}

\numberwithin{equation}{section}

\renewcommand{\theequation}{\arabic{section}.\arabic{equation}}

\def\IR{\mathbb{R}}
\def\IN{\mathbb{N}}

\newcommand{\be}{\begin{equation}}
\newcommand{\ee}{\end{equation}}
\newcommand{\bea}{\begin{eqnarray}}
\newcommand{\eea}{\end{eqnarray}}

\def\rm{\mathrm}
\newcommand{\Db}[1]{$\overline{\mathrm{D}#1}$}

\def\ti{\widetilde}

\global\arraycolsep=1pt
\def\bea{\begin{eqnarray}}
\def\eea{\end{eqnarray}}
\def\be{\begin{equation}}
\def\ee{\end{equation}}
\def\bem{\begin{multline}}
\def\eem{\end{multline}}

\def\QN{{\mathit{Q}}}

\newcommand{\Scal}[1]{\Bigl ({#1} \Bigr )}
\newcommand{\scal}[1]{\bigl ({#1} \bigr )}
\newcommand{\CR}{\nonumber \\*}

\newcommand{\trace}{\hbox {Tr}~}

\DeclareMathAlphabet{\mathpzc}{OT1}{pzc}{m}{it}

\DeclareMathOperator{\ad}{ad}

\newcommand{\ord}[1]{{\scriptscriptstyle (#1)}}

\DeclareMathAlphabet{\mathpzc}{OT1}{pzc}{m}{it}

\newcommand{\IC}{\mathbb{C}}

\def\deux{{\mathpzc{2}}}

\def\quatre{{\mathpzc{4}}}

\newcommand{\stfrac}[2]{{\textstyle \frac{#1}{#2}}}

\def\DSOVIII#1#2#3#4{{\tiny $ {   \biggl[ \begin{array}{ccc}  &&\mathfrak{#2}  \vspace{ -1.5mm} \\  \mathfrak{#1}\hspace{-0.6mm} &\mathfrak{#4} \hspace{-0.9mm}&\vspace{-1.5mm}\\ && \mathfrak{#3}  \end{array}\biggr] }$}}
\def\DDSOVIII#1#2#3#4{{\tiny $ {   \biggl[ \begin{array}{ccc}  &&{#2}  \vspace{ -1.5mm} \\  {#1}\hspace{-0.6mm} &{#4} \hspace{-0.9mm}&\vspace{-1.5mm}\\ && {#3}  \end{array}\biggr] }$}}

\def\ie{{\it i.e.}\ }
\def\eg{{\it e.g.}\ }

\def\pA{{\scriptscriptstyle A}}
\def\pB{{\scriptscriptstyle B}}

\def\e{\boldsymbol{e}}

\def\asym{{\scriptscriptstyle 0}}

\def\g{\mathfrak{g}}

\def\V{{\mathcal{V}}}

\def\KK{{K\hspace{-1mm}K}}

\def\gl{\mathfrak{gl}}
\def\sl{\mathfrak{sl}}
\def\so{\mathfrak{so}}

\def\e{\mathfrak{e}}

\def\DJo{$\;$\kern-.4em \hbox{D\kern-.8em\raise.15ex\hbox{--}\kern.35em okovi\'c}}

\def\N{\mathcal{N}}
\newcommand{\eprint}[1]{{\href{http://arxiv.org/abs/#1}{\texttt{[#1}]}}}
\newcommand{\eprintN}[1]{{\href{http://arxiv.org/abs/#1}{\texttt{#1 [hep-th]}}}}

\begin{document}

\begin{titlepage}

\begin{flushright}
CPHT-RR050.0611\\
AEI-2011-038
\end{flushright}

\bigskip
\bigskip
\centerline{\Large \bf Interacting non-BPS black holes}
\centerline{\Large \bf }
\bigskip
\bigskip
\centerline{{\bf Guillaume Bossard$^1$ and Cl\'{e}ment Ruef$^{\,2}$}}
\bigskip
\centerline{$^1$ Centre de Physique Th\'eorique, Ecole Polytechnique, CNRS}
\centerline{91128 Palaiseau cedex, France}
\bigskip
\centerline{$^2$ Max Planck Institute for Gravitation, Albert Einstein Institute}
\centerline{Am M\"uhlenberg 1, 14476 Golm, Germany}
\bigskip
\centerline{{ bossard@cpht.polytechnique.fr,~clement.ruef@aei.mpg.de, } }
\bigskip
\bigskip

\begin{abstract}

We explain how to exploit systematically the structure of nilpotent orbits to obtain a solvable system of equations describing extremal solutions of (super-)gravity theories, {\it i.e.} systems that can be solved in a linear way. We present the procedure in the case of the STU model, where we show that all extremal solutions with a flat three-dimensional base are fully described with the help of three different nilpotent orbits:  the BPS, the almost-BPS and the composite non-BPS. The latter describes a new class of solutions for which the orientation of half of the constituent branes have been inverted with respect to the BPS one, such that all the centres are intrinsically non-BPS, and interact with each others. We finally recover explicitly the ensemble of the almost-BPS solutions in our formalism and present an explicit two-centre solution of the new class.

\end{abstract}

\end{titlepage}


\section{Introduction}

One of the great success of string theory has been to be able to provide a statistical interpretation to the Bekenstein--Hawking entropy of BPS black holes through the counting of D-branes in the weakly coupled regime \cite{Strominger:1996sh}. The validity of the computation is ensured by supersymmetry, but it has nevertheless been proposed that this property could generalise to non-BPS extremal black holes \cite{Emparan,DabhoSen,Dimitru}. The classification of supersymmetric composite black hole solutions has permitted to understand the mismatch between the enumeration of spherically symmetric BPS black holes and the counting of BPS states within weakly coupled string theory \cite{Denef}. Understanding the space of states associated to extremal black holes therefore clearly requires to have a global understanding of composite extremal black hole solutions. 
In the recent years, a lot of techniques developed in the context of supersymmetric solutions have been adapted to the case of non-BPS extremal ones (see \cite{Ceresole:2007wx}--\cite{Galli:2011fq} for part of the literature). Indeed, it has been understood that a lot of features are in fact intrinsically more related to extremality than supersymmetry. 
In particular, the underlying system of equations is then {\it solvable}, which means that it can be solved in a linear fashion \cite{Bena:2009fi}. Such a linear structure is the key point for solving explicitly the equations, and this remark permitted to construct a lot of new non-BPS solutions \cite{Bena:2009ev,Bena:2009en,Bena:2009fi}. However, it has up to now remain unclear where such a solvable system was coming from, and thus how to generalise the approach to find other solvable systems. The aim of the present paper is to address this issue. 

Since more than twenty years now, it is well-known that the stationary solutions of supergravity theories coupled to abelian vector fields and scalar fields parametrizing a symmetric space $G_\quatre / K_\quatre$ are described by a non-linear sigma model coupled to Euclidean three-dimensional gravity, which scalar field $\V$ parametrize a symmetric space $G/K^*$ \cite{Breitenlohner:1987dg}. In the cases of interest (such has Kaluza--Klein supersymmetric theories without hyper-multiplets), $G$ is a simple group and $K^*$ is a non-compact real form of its maximal compact subgroup. The scalar momentum $P$ is defined as the component of the Maurer--Cartan form $\V^{-1} d \V$ in the coset component $\mathfrak{p} \cong \mathfrak{g} \ominus \mathfrak{k}^*$ of the Lie algebra. Solutions describing spherically symmetric black holes are then determined by the associated Noether charge in the Lie algebra $\mathfrak{g}$ of $G$, and can therefore be classified in terms of $G$-orbits \cite{Breitenlohner:1987dg,Josef}. In the extremal limit, the Noether charge is {\it nilpotent} and the spherically symmetric extremal black hole solutions are classified in terms of a sub-class of nilpotent orbits of $G$ in $\mathfrak{g}$ \cite{Bossard:2009at,Bossard:2009my}. It is shown that $P$ then lies in the same nilpotent orbit, and more precisely in its intersection with the coset component $\mathfrak{p}$. It has been exhibited that the first order system of differential equations defining the composite BPS black hole solutions can itself be derived from the property that the scalar momentum $P$ lies in the nilpotent orbit associated to spherically symmetric BPS black holes \cite{BossardBPS}. We show in this paper that this property extends to all extremal composite black hole solutions with a flat three-dimensional base metric,\footnote{Excluding the extremal Kerr solutions and composite generalisations thereof.} without assuming any spherical symmetry, nor even axisymmetry. In other words, they are solutions of a solvable system of differential equations, which is defined by the nilpotent orbit in which $P$ is constrained to lie in. We will be able to obtain {\it all the possible solvable systems} of a given theory from the study of its nilpotent orbits, and also to explore {\it in a systematic way} the structure of the solutions space. 
%
%
In the present paper, for the sake of clarity, we only study the case of the STU model, \ie $\N=2$ supergravity in four-dimensions coupled to three vector multiplets with a cubic prepotential. The ideas explained here are nevertheless more general, and we intend to present the more general case of $\N=8$ supergravity in an upcoming paper. The generalisation to any $\N=2$ supergravity coupled to vector multiplets associated to a very special K\"{a}hler geometry is straightforward. For the STU model, we show that all solutions can be described with only three different nilpotent orbits, leading to three inequivalent solvable system of differential equations. The first one is the Denef system of equations for BPS solutions \cite{Denef,Bates} and the second is the almost-BPS system \cite{Goldstein:2008fq}. We obtain the latter from the nilpotent orbit approach, and show that it allows us to recover all the previously known solutions of that class. The third one is new, and we call it the composite non-BPS system. In terms of type IIA supergravity, one can understand the almost-BPS system as a system of floating D-branes for which the constituent branes of one type have an inverted orientation, \eg \Db{6}-D4-D2-D0. Similarly, the composite non-BPS system is associated to a system of D-branes for which the constituent branes of two types have their orientation inverted, \eg \Db{6}-\Db{4}-D2-D0. We present here the composite non-BPS system of equations, but, again for the sake of simplicity, only present one two-centres solution of this class. This example will be enough to exhibit key properties of the new class, in particular the fact that the centres are interacting, and to show that more general solutions exist with an arbitrary number of centres.

\begin{figure} 
\begin{psfrags}
	\psfrag{t1}{\tiny $\frac{1}{2}$-BPS}
	\psfrag{t2}{\tiny $\frac{1}{4}$-BPS}
	\psfrag{t3}{\tiny $\frac{1}{8}$-BPS, $S=0$}
	\psfrag{t4}{\tiny \com{$\frac{1}{8}$-BPS}}
	\psfrag{t5}{\tiny Physical non-BPS}
	\psfrag{t6}{\tiny }
	\psfrag{t7}{\tiny \com{composite non-BPS}}
	\psfrag{t8}{\tiny \com{almost-BPS}}
	\psfrag{t9}{\tiny {\color{red} Regularity line}}
\includegraphics[width=\linewidth]{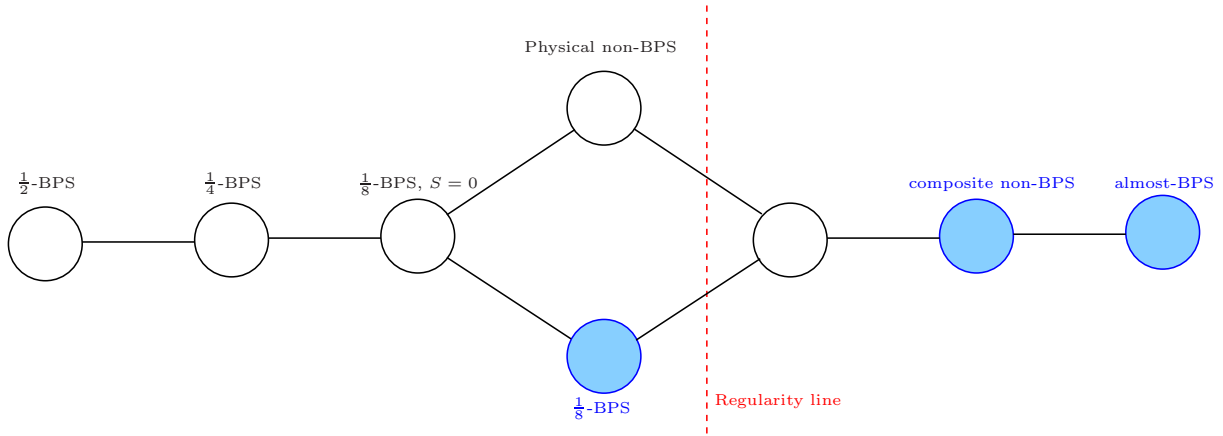}
\end{psfrags}
\caption{\small This figure depicts a simplified version of the Hasse diagram of the nilpotent orbits of $SO_\asym(4,4)$, where the orbits have been identified when $ E_{8(8)}$ conjugate in $\e_{8(8)}$. It describes the topology of the space of nilpotent elements in $\so(4,4)$, such that two nilpotent orbits are connected by a line if the left orbit lies in the topological closure of the right one, see \cite{Collingwood} for details. The lower rank orbits, on the left, corresponds to $1/2$, $1/4$ and $1/8$-BPS solutions of $\N=8$ supergravity with a null entropy. Then, two orbits have the same rank, they are the $1/8$-BPS and the physical non-BPS orbit associated to the regular extremal single-centre solutions. On the right side of the regularity line (dotted red line), the rank of the orbits are too high to describe regular single-centre solutions, but still allows for multi-centre configurations. In this paper, we study in details the three blue shaded orbits of this diagram. All the extremal solutions can be obtained from them as subcases.} \label{Hassediag1}
\end{figure}

An important point in understanding extremal solutions is the following. A generic single-centre solution of both the almost-BPS and the composite non-BPS system will be singular, and thus not physical. In order to be regular, the momentum $P$ must fall down at any black hole horizon, either in the BPS or the so-called physical non-BPS orbit, which are associated to extremal single-centre solutions. However, regularity allows $P$ to lie in more general higher rank nilpotent orbits on a generic point of the base space. This is schematically depicted in Figure \ref{Hassediag1}. It is crucial to understand that this is precisely the possibility for $P$ to lie in higher rank orbits, where a single-centre solution would not be regular, that permits to obtain interacting non-BPS multi-centres solutions. Almost-BPS solutions can describe multi-centre interacting solutions \cite{Bena:2009ev,Bena:2009en}, but no interaction can take place between two non-BPS centres. It is of importance that the composite non-BPS multi-centre black holes that we exhibit in the present paper have interactions between genuinely non-BPS centres. This is an important step toward the understanding of the structure of the moduli space of non-BPS multi-centre solutions. 

Note also that the existence of such solutions is not in contradiction with the previously obtained conclusion that there were no composite non-BPS solutions associated to the first order system of differential equations describing non-BPS single-centre solutions \cite{Gaiotto:2007ag}. Indeed we find that the most general regular solution of the `physical' non-BPS system (associated to the `physical' non-BPS nilpotent orbit) are single-centre solutions with a possible bounded angular momentum, whereas the composite non-BPS solutions only exist in a more general system, the composite non-BPS system. One main difference with the BPS solutions is, whereas for the latter the ADM mass and the flow of the scalar fields in the asymptotic region of a supersymmetric space-time are entirely determined by the asymptotic central charges $Z(q,p), Z_i(q,p)$, they also depend on the specific structure of the interior space-time for a composite non-BPS solution. This can be interpreted within the attractor mechanism by a lifting of the flat directions normally associated to single-centre non-BPS black holes. In the presence of interactions, the flow of the scalar fields in the asymptotic region is indeed governed by the `auxiliary field' dependent `fake superpotential' described in \cite{Ceresole:2009vp}, such that instead of being determined by extremizing the `fake superpotential', the `auxiliary fields' associated to the flat directions are determined in function of the specific structure of the interior space-time.

The paper is organised as follows. In section 2 we present, from a purely group theoretical point of view, how nilpotent orbits generically define solvable systems, and then focus on the case of the STU model where we obtain three solvable systems that encode all extremal solutions of the STU model with a flat base. We restrict the arguments based on group theory to this section, and the reader not interested in the mathematical details can consistently skip this part, its results being recalled when needed in the rest of the paper. Section 3 is devoted to set up the theory and the conventions, in particular to relate the three-dimensional and four-dimensional quantities. In this section we also provide the expression of the ADM mass as a function of the asymptotic central charges and phase parameters determined by the specific configuration of the interior solution. In section 4, we choose a duality frame to show how, from our point of view, we reobtain all known almost-BPS solutions, while section 5 is devoted to the new composite non-BPS system. In this two cases, we also show that the Ehlers rotation, which is part of the three-dimensional duality group, but not of the four-dimensional one, does not give any new solutions. Finally, we construct and analyse a complete two-centre solution of the composite non-BPS class in section 6. Some details of the conventions are relegated to an appendix.

\section{Solvable subalgebras and equations of motion}

\subsection{General solvable subalgebra}
\label{GSS} 

One interesting feature of the almost-BPS system of differential equations \cite{Goldstein:2008fq,Bena:2009ev}, is that, although not free, it admits a graded structure that allows to obtain each function as the solution of a free Laplace equation with a source term determined as a non-linear function of the lower grade functions. This renders the system exactly solvable, \ie it amounts to recursively solve linear Laplace equations with known non-linear source terms. We explain in this section how this behaviour is in fact generic for extremal systems. In terms of the non-linear sigma model over $G/K^*$ coupled to Euclidean gravity in three dimensions, this graded structure originates naturally from the graded structure of a solvable subalgebra.

\medskip

Consider a solvable subalgebra $\mathfrak{n} \subset \g$, which satisfies by definition that 
\be \label{nilpad}
 \exists \, n \in \IN \,, \quad  \ad^{\; n}_\mathfrak{n} \mathfrak{n}  \cong \{ 0 \} \, ,
\ee
where 
\be  
\ad_\mathfrak{n} {\mathfrak{n}}\cong [ \mathfrak{n} , \mathfrak{n} ] 
\ee
as a set, and the power $n$  defines the number of commutators. $\mathfrak{n} $ admits a grading
\be
 \mathfrak{n} = \sum_{p=1}^n \mathfrak{n}^\ord{p}\; ,
\ee
such that
\be
  \mathfrak{n}^\ord{p} \cong \ad^{\; p-1}_\mathfrak{n} \mathfrak{n} \setminus  \ad^{\; p}_\mathfrak{n} \mathfrak{n} \,\; .
\ee
We will consider that this grading is consistent with the involution defining the subalgebra $\mathfrak{k}^*$
\be
 \mathfrak{g} \cong \mathfrak{k}^* \oplus {\mathfrak{p}} 
\ee
associated to $K^*$, such that each component decomposes accordingly into
\be 
\mathfrak{n}^\ord{p} \cong \mathfrak{k}^\ord{p} \oplus \mathfrak{p}^\ord{p} \,.
\ee  
 
Now, consider the Ansatz for the scalar field
\be \label{VeqexpL}
\V = \exp(-L) 
\ee
defined such that $L$ is a function of the three-dimensional base 
 $M_{\scriptscriptstyle 3}$ valued in $\mathfrak{n} \cap \mathfrak{p}$
\be \label{NinterP}
L : M_{\scriptscriptstyle 3}  \rightarrow \mathfrak{n} \cap \mathfrak{p} \,.
\ee
$L$ is in $\mathfrak{n}$ because we assume it to be nilpotent, and it can be chosen in $\mathfrak{p}$ by fixing the coset representative  $\V$  in $G/K^*$ to be odd with respect to the involution defining $K^*$.\footnote{Note that this choice of parametrization of the coset space $G/K^*$ is not globally defined, because $G/K^*$ is not topologically trivial as opposed to $G/K$. Nevertheless it is well defined on a dense subspace and the singular loci correspond to singularities of the four-dimensional metric and can be disregarded.}

 One straightforwardly obtains that the components of the Maurer--Cartan form 
\be 
\V^{-1} d \V = P + B \; , \qquad P \in \mathfrak{p} \; , \quad B \in \mathfrak{k}^* 
\ee
are 
\be 
P = -\sum_{k\ge0} \frac{1}{(2k+1)!} \ad_L^{\; \; 2k} d L  \; , \qquad B = -\sum_{k\ge 0} \frac{1}{(2k+2)!} \ad_L^{\;\; 2k+1} d L  \; , 
\ee
and lie respectively in 
\be 
 P \in \mathfrak{n} \cap \mathfrak{p} \cong \sum_{p=1}^n \mathfrak{p}^\ord{p}\; , \qquad B \in   \mathfrak{n} \cap \mathfrak{k}^* \cong \sum_{p=1}^n \mathfrak{k}^\ord{p} \,. \ee
Note that both sums are finite since $\ad^{\; \; n}_L dL =0$. By property of the solvable algebra 
\be \label{P2eq0}
 \trace P_\mu P_\nu = 0 \,. 
\ee

\paragraph{Equations of motion.}

The equations of motion of (super-)gravity theories coupled to abelian vector fields in four dimensions, reduce for stationary solutions to the following equations on the three-dimensional Riemannian space $M_{\scriptscriptstyle 3} $
\bea
 \trace P_\mu P_\nu &=& R_{\mu\nu} \quad  , \quad \label{eom1} \\ \label{eom2}
 d \star P +  [ B , \star P ] &=& 0 \quad .
\eea
So \eqref{P2eq0} translates into $R_{\mu\nu}=0$, and the base three-dimensional metric is flat $\gamma_{\mu\nu}  = \delta_{\mu\nu}$. Therefore the base manifold is simply $\mathds{R}^3$, although it will be more precise to define it as the punctured $\mathds{R}^3$
\be 
M_{\scriptscriptstyle 3} \cong \mathds{R}^3 \setminus \{ x_{\scriptscriptstyle A}|_{A=1}^k \} \quad , 
\ee
where the removed points $x_{\scriptscriptstyle A}$ are the poles of the function $L$. To look at the second equation of motion \eqref{eom2}, it is useful to define $L^\ord{p} \in \mathfrak{p}^\ord{p}$  such that 
\be 
 L \equiv \sum_{p=1}^n L^\ord{p} \,.
\ee
One can then develop \eqref{eom2} according to the grading such that 
\bea \label{galgradedeq}
 d\star d L^\ord{1} &=& 0 \CR
 d\star d L^\ord{2} &=& 0 \CR
 d \star d L^\ord{3} &=& - \frac{2}{3} \bigl[ d L^\ord{1} , [ L^\ord{1} , \star d L^\ord{1}] \bigr] \CR
 d \star d L^\ord{4} &=&   - \frac{2}{3} \bigl[ d L^\ord{1} , [ L^\ord{1} , \star d L^\ord{2}] \bigr]  - \frac{2}{3} \bigl[ d L^\ord{1} , [ L^\ord{2} , \star d L^\ord{1}] \bigr]  - \frac{2}{3} \bigl[ d L^\ord{2} , [ L^\ord{1} , \star d L^\ord{1}] \bigr] \CR
  d \star d L^\ord{5} &=&   \frac{2}{45} \bigl[ dL^\ord{1} , [ L^\ord{1} , [ L^\ord{1} , [ L^\ord{1} , \star d L^\ord{1} ]]] \bigr] + \frac{8}{45} \bigl[ [ L^\ord{1} , d L^\ord{1}] , [ L^\ord{1} , [ L^\ord{1} , \star d L^\ord{1} ]] \bigr]  \CR
   && - \frac{2}{3} \bigl[ d L^\ord{1} , [ L^\ord{2} , \star d L^\ord{2}] \bigr]  - \frac{2}{3} \bigl[ d L^\ord{2} , [ L^\ord{1} , \star d L^\ord{2}] \bigr]  - \frac{2}{3} \bigl[ d L^\ord{2} , [ L^\ord{2} , \star d L^\ord{1}] \bigr] \CR
   &&  - \frac{2}{3} \bigl[ d L^\ord{1} , [ L^\ord{1} , \star d L^\ord{3}] \bigr]  - \frac{2}{3} \bigl[ d L^\ord{1} , [ L^\ord{3} , \star d L^\ord{1}] \bigr]  - \frac{2}{3} \bigl[ d L^\ord{3} , [ L^\ord{1} , \star d L^\ord{1}] \bigr] \CR
   d \star d L^\ord{6} &=& \; \dots \label{GradedSystem}
 \eea
and so on and so forth, such that each component $L^\ord{p}$ of $L$ can be obtained as the solution of a Laplace equation with a source depending of the function $L^\ord{q}$ of lower grade $q<p-1$. In other words, one can solve these equations {\it linearly}. It should be clear at this point that this is a generic property of solutions described by {\it any} solvable subalgebra. Note that because of the nilpotency \eqref{nilpad}, the system \eqref{galgradedeq} stops for a finite degree $n$.

The metric and electromagnetic fields defined in four (respectively five) dimensions are defined in function of scalars determined as algebraic functions of the component functions of $L$, and $1$-forms determined as components of the $\mathfrak{n}$ valued $1$-form $W$ dual to the scalar fields
\be 
 d W \equiv \star \V P \V^{-1} = - \sum_{k =0}^{n-1} \frac{(-2)^k}{(k+1)!} \ad_L^{\; \; k}  \star d L \quad , 
\ee 
which is well defined on a $U(1)^{\times \mbox{dim}[\mathfrak{n}]}$ bundle over $M_{\scriptscriptstyle 3}$ since the equations of motion imply 
\be \label{ConservLaw}
 d  \star \V P \V^{-1} = 0 \,.   
\ee
This also provides a Noether charge $\QN \in \mathfrak{n}$ associated to any $2$-cycle $\Sigma$ of the three-dimensional base manifold
\be
 \QN_{|\Sigma} \equiv \frac{1}{4\pi} \int_{\Sigma} \star \V P \V^{-1} \,.
\ee

\paragraph{Regularity.}

Because we will be interested in composite black hole solutions, we can assume that the cycles are characterised by the black holes they are surrounding. In particular, since we will study extremal black holes, they will be characterised by poles in the function $L$ at point  $x_{\scriptscriptstyle A}$ located at the horizon of the black holes. By definition of the system, the Noether charge will lie in $\mathfrak{n}$, and therefore will always be nilpotent. The Noether charge associated to a cycle surrounding one single black hole is characterised by the pole of $ \star \V P \V^{-1}$ at $x = x_{\scriptscriptstyle A}$. The geometry of the horizon is only modified by the existence of the other centres by subleading corrections, and in order for the solution to be regular, the associated Noether charge must satisfy at least the constraint that it does satisfy for the single-centre solution to be regular. A single-centre extremal solution is regular if and only if it can be obtained as the extremal limit of a regular non-extremal solution. This requires for instance that $\QN_{\scriptscriptstyle A}$ is regular in the boundary of the $ \mathds{R}_+^*\times K^*$ orbit of a Schwarzschild charge $\QN_{\scriptscriptstyle \rm {Sch}} = M {\bf H}$ \cite{Breitenlohner:1987dg},
\be
 \QN_{\scriptscriptstyle A} \in \partial\bigl(   \mathds{R}_+^*\times K^* \cdot {\bf H}  \bigr) \quad , \qquad \mbox{dim}\bigl[K^* \cdot \QN_{\scriptscriptstyle A}\bigr] = \mbox{dim}\bigl[  K^* \cdot {\bf H}  \bigr]  \; , 
\ee
where $({\bf H}, {\bf E}, {\bf F})$ is the $\sl_2$ triple inside $\g$ which defines the pure gravity truncation. For instance, a pure gravity solution would read $\V = \exp(\sigma {\bf E}) \exp( U {\bf H}) $ and the Schwarzschild solution would for example be defined such that 
\be \V = \exp\left( \ln \frac{r-M}{r+M} \; {\bf H} \right) \; . \ee
We define also $k_\g$ such that  $2 k_{\g} \equiv \trace {\bf H}^2 $. The absence of Dirac--Misner string singularities requires moreover that the Kaluza--Klein vector $\omega$ -- the four-dimensional angular momentum vector -- is a globally defined $1$-form on $M_{\scriptscriptstyle 3}$. The latter will always be defined as the specific component of $W$
\be
  \omega = \frac{1}{k_{\g}} \trace {\bf E}\,  W \; ,
\ee
${\bf E}$ being still the positive nilpotent element of the pure gravity $\sl_2$ triple inside $\g$. The absence of Dirac--Misner string singularities requires therefore
\be
 \trace {\bf E} \,  \QN_{\scriptscriptstyle A} = 0 \; , 
\ee
for all centres. 

On any cycle $\Sigma(I)$ surrounding centres $x_{\scriptscriptstyle A}$ for $A\in I$, the Noether charge is 
\be
 \QN_{|\Sigma(I)} = \sum_{A\in I} \QN_{\scriptscriptstyle A} \,.
\ee 
A crucial point to be understood, is that although each $\QN_{\scriptscriptstyle A}$ is required, by regularity, to be nilpotent of order 3, The sum of the charges corresponding to different centres, and in particular the asymptotic charge of the solution, do not need to be in such a `regular' orbit. Thus, the charge associated to a generic cycle can in general be nilpotent of higher degree. Therefore we expect the general solution to admit a total charge
\be
 \QN = \sum_A \QN_{\scriptscriptstyle A} \; , 
\ee
which could be any nilpotent element of $\mathfrak{g}$, provided the corresponding $G$-orbit admits an intersection with the coset component $\mathfrak{p}$.  For such a solution, $P$ will be valued in a higher order nilpotent orbit on $M_{\scriptscriptstyle 3}$, although its poles at $x_{\scriptscriptstyle A}$ will lie in $K^*$ orbits associated to regular single-centre solutions. This is depicted in figure \ref{Hassediag1}. As we will see, this remark is the key point to obtain interacting non-BPS multi-centre solutions.

\subsection{Parametrizing nilpotent orbits.}

We are therefore interested in classifying the solvable subalgebras $\mathfrak{n}$ in $\g$ in function of the nilpotent orbits their generic elements lie in (see  \cite{Collingwood} for a pedagogical introduction to nilpotent orbits). The consistency of the grading with the involution defining $K^*$ implies that it can always be defined from a particular semi-simple element $h$ (\ie which is diagonalizable in a faithful complex representation) in $\mathfrak{k}^*$ such that for any element ${\rm x}^\ord{p} \in \mathfrak{n}^\ord{p}$,
\be
 [ h , {\rm x}^\ord{p} ] = 2 p \, {\rm x}^\ord{p} \,.\label{evenGrad}
\ee 
It appears that the nilpotent orbits are themselves classified in terms of such semi-simple generator by mean of the so-called normal triples. Here the factor of two implies that the corresponding nilpotent orbit is even, and we will always assume the nilpotent orbits to be even.\footnote{This does not assume any lost of generality because any nilpotent element of an odd nilpotent orbit always lies in a solvable algebra associated to an even nilpotent orbit.} The conjugacy class $G_{\mathds{C}}\cdot e$ with respect to the complex Lie group $G_{\mathds{C}}$ of a nilpotent element $e\in \g_\mathds{C}$ is entirely determined by the conjugacy class $G_{\mathds{C}} \cdot h$ of a semi-simple element defining a standard $\sl_2$ triple $(h,e,f)$ verifying
\be
 [ h , e] = 2e \; , \qquad [ e , f ] = h \; , \qquad [ h , f ] = - 2f \,.
\ee
By construction, $e$ is nilpotent, but, from \eqref{NinterP}, we will also require it to lie in the coset component $\mathfrak{p}$. In this case, $h$ can always be chosen in the $\mathfrak{k}^*$ subalgebra of $\g$, $h \in \mathfrak{k}^*$, and the triple is then called a normal triple \cite{Collingwood}. It will therefore be natural for us to classify solvable subalgebras in terms of conjugacy classes of semi-simple elements characterizing complex nilpotent orbits under the subgroup $K_{\IC} \subset G_{\IC}$, $K_{\IC} \cdot e \subset  \mathfrak{p}_{\IC}$. The classification of real orbits requires more work and is of no use for our purpose, so we will not discuss it in this paper. Indeed, we would like to straight that also the function of interest $L$ is clearly real, the first order system is only characterised by the complex nilpotent orbit it lies in, and it will appear in regular solutions that $P$ `jumps' from one real $K^*$-orbit from one another within the same complex orbit. To summarise, a nilpotent element $e \in  \mathfrak{n} \cap \mathfrak{p}$ can be associated a normal $\sl_2$ triple $(h,e,f)$, such that its $K_{\IC}$ conjugacy class is determined by the $K_{\IC}$ conjugacy class of the corresponding semi-simple element $h \in \mathfrak{k}^*$.

Consider the nilpotent element $e \in \mathfrak{p}$ of a normal triple $(h,e,f)$ such that $h\in \mathfrak{k}^*$ defines an even grading (\ref{evenGrad}) of $\g$. By definition, the positive grade component $\sum_{k>0}  \mathfrak{g}^\ord{k}$ defines a solvable algebra containing $e$, and the maximal solvable algebra containing $\sum_{k>0}  \mathfrak{g}^\ord{k}$ is necessarily a subalgebra of $\sum_{k\ge0}  \mathfrak{g}^\ord{k}$. Here we will assume that all the maximal solvable subalgebras are the strict positive grade component  $\sum_{k>0}  \mathfrak{g}^\ord{k}$ associated to specific normal triples. This is true for the case of $\so(4,4)$ which we will study in this paper, as can be shown by inspection, but we will explain in a forthcoming work that this does not exhaust the set of maximal solvable subalgebras for $\e_{8(8)}$.  

\bigskip

Now considering a solvable algebra $\mathfrak{n} \cong \sum_{k>0}  \mathfrak{g}^\ord{k}$ associated to a normal tripe $(h,e,f)$, we would like to parametrize the set of inequivalent embeddings of this algebra inside $\g$. The action of $K^*$ clearly preserves the graded system of differential equations, and acts transitively on the set of embeddings of the normal triple. Physically, the inequivalent embeddings of strict normal triples correspond to the inequivalent parametrizations of the four-dimensional fields in terms of the set of functions satisfying the graded system of differential equations, or in other words, to the inequivalent duality frames  the solution can be written in. Define $\KK\subset K^*$ the maximal compact subgroup of $K^*$ and $K^+(h)\subset K^*$ the parabolic subgroup generated by the Lie algebra $\sum_{k\ge0}  \mathfrak{k}^\ord{k}$ associated to $h$. By Iwasawa decomposition, any element $u \in K^*$ is the product of an element $u_0 \in \KK$ and an element $u_+ \in K^+(h)$, $u=u_0 \cdot u_+$. Then, by Sekiguchi lemma \cite{Sekiguchi}, a normal triple $(h,e,f)$ is conjugate under $K^*$ to a strict normal triple $(\tilde{h},\tilde{e},\tilde{f})$ such that $\tilde{h}$ is odd with respect to the Cartan involution, \ie $\tilde{h}$ is Hermitian in a matrix representation. But, because $u_0$ is invariant with respect to the Cartan involution, any normal triple $(h,e,f)$ is conjugate under $K^+(h)$ to a strict normal triple $(\tilde{h},\tilde{e},\tilde{f})$. Because $K^+(h)$ preserves $\mathfrak{n}$ by definition, the solvable algebra can always be associated to a strict normal triple $(h,e,f)$, and so the generator $h$ defining the nilpotent algebra can always be chosen to be odd with respect to the  Cartan involution. It follows that the $K^*$ orbit of inequivalent embeddings of $\mathfrak{n} \subset \g$ is isomorphically parametrized by the $\KK$-orbit of the corresponding $h$. Note that for the physical models, $\KK$ is always the product of the Ehlers $U(1)$ and the maximal compact subgroup $K_\quatre$ of the four-dimensional duality group $G_\quatre$. We have thus shown that the inequivalent duality frames corresponding to a solvable system associated to a nilpotent orbit are parametrized by the $U(1) \times K_\quatre$ orbit of the generator $h$ defining a strict normal triple. In fact we will see that the action of the Ehlers $U(1)$ is physically irrelevant, because the inequivalent duality frames it permits to define do not support any regular black hole type solutions.  The set of `physically relevant' inequivalent duality frames associated to a solvable system of differential equation is therefore parametrized by the $K_\quatre$-orbit of an associated strict normal triple generator $h$.

\bigskip

%
%
In this paper we will restrict ourselves to the example of the $STU$ model, for which the three-dimensional scalars parametrize the `pseudo-quaternionic' symmetric space 
\be
 G/K^* = SO(4,4) / \scal{ SL(2) \times_{\mathds{Z}_2} SL(2) \times SL(2) \times_{\mathds{Z}_2} SL(2) } \; , 
\ee 
and so $h \in \bigoplus_{\Lambda= 0}^3 \sl_2^\ord{\Lambda}$.  Choosing a particular $F_{\Lambda} , H_{\Lambda} , E_{\Lambda}$ basis for each $\sl_2^\ord{\Lambda}$, the generator $h$ determining a nilpotent orbit is parametrized by four positive half integers $b^\Lambda$ such that 
\be
 h = \sum_\Lambda b^\Lambda H_\Lambda \,.
\ee
This parametrization is the key point to obtain the graded system of equations, as we will see in the following. Because a nilpotent orbit of $K^*$ is always defined as a Lagrangian submanifold of a nilpotent orbit of $G$, it will be relevant to define the conjugacy class of $h \in \mathfrak{g}$ by its so-called weighted Dynkin diagram \DDSOVIII{\beta_1}{\beta_2}{\beta_3}{\beta_0}, which evaluates a dominant representative of $G \cdot h$ in a Cartan subalgebra on the corresponding simple roots of $\so(4,4)$. It is proved that $\beta_\Lambda \in \{ 0,1,2\}$ ($\beta_\Lambda \in \{ 0,2\}$ for even orbits) \cite{Collingwood}. 

\bigskip

The parameters $b^\Lambda$ associated to a Noether charge $\QN$ corresponding to one centre can be characterised by the central charges at the horizon as 
\be
 b^\Lambda(\V_*{}^{-1}  \QN\V_* ) = \frac{2}{\sqrt[4]{|I_4(q,p)|}}  |Z_{\Lambda\, *} (q,p)|  \; , 
\ee
where $q_\Lambda,p^\Lambda$ are the electromagnetic charges, and $I_4(q,p)$ the corresponding quartic invariant, 
\be \label{quarticinvariant}
 I_4(q,p) = 4 p^0 q_1 q_2 q_3  -4 q_0 p^1 p^2 p^3  + 4 \sum_i p^{i+1} p^{i+2} q_{i+1} q_{i+2} - \biggl( p^0 q_0 + \sum_i p^i q_i \biggr)^2   \,.
\ee
$Z_{0\, *}(q,p) \equiv Z_{*}(q,p) $ is the central charge at the horizon and $ Z_{i\, *}(q,p) $ its K\"ahler derivatives in tangent frame.\footnote{ \ie $Z_i(p,q) \equiv - ( t^i - \bar t^i )  e^{- \frac{\mathcal{K}}{2}} \partial_i  \bigl( e^{ \frac{\mathcal{K}}{2}} Z(q,p) \bigr) $.}  $\V_*{}^{-1}  \QN\V_* $ describes the pole in $P\in  \mathfrak{p}$ at the centre as a residue, where $\V_*$ is the value of the coset representative at the horizon. Although $\V$ diverges at the pole, $\V_*{}^{-1}  \QN\V_* $ is well defined at the horizon because the singular component of $L$ commutes with $\QN$ by definition of the charge. 

The BPS black holes admit a nilpotent Noether charge such that $b^\Lambda = (2,0,0,0)$, the non-BPS black holes with a vanishing central charge at the horizon correspond to the permutations $b^\Lambda = (0,2,0,0),\; (0,0,2,0),\;  (0,0,0,2)$ and will not be discussed,\footnote{They would be BPS in a different truncation of maximal supergravity.} and the `physical' non-BPS black holes with a non-vanishing central charge at the horizon admit a Noether charge such that $b^\Lambda = (1,1,1,1)$.

\subsection{Three solvable subalgebras describing all solutions}

There are three solvable subalgebras that contain all the others by permutations of the $b^\Lambda$'s or as subalgebras, and so we will focus on them.\footnote{For a discussion of the nilpotent orbits of $SO_\asym(4,4)$ see \cite{DokovicSO,BossardW}.} These three algebras have the same dimension, and describe three different families of solutions, each of them being characterised by 8 harmonic functions. The first is the well-known BPS system, which can be interpreted in type IIA as describing D6-D4-D2-D0 bound states. The second is the almost-BPS one \cite{Goldstein:2008fq,Bena:2009ev} and describes type IIA bound states where the constituent branes of one type have their orientation inverted, as \eg \Db{6}-D4-D2-D0, while the last one corresponds to type IIA configurations in which the orientation of half of the branes is inverted, as \eg \Db{6}-\Db{4}-D2-D0.

\subsubsection{The BPS algebra}

Let us start by the well known example of the BPS Denef system \cite{Denef,Bates}. In that case the pertinent subalgebra is determined by $b^\Lambda = (2,0,0,0)$, {\it i.e.} $h =2 H_0$. $h\in \so(4,4)$ is then characterised by the weighted Dynkin diagram \DSOVIII0002 and defines the following graded decomposition of $\so(4,4)$
\be
 \so(4,4) \cong {\bf 1}^\ord{-2} \oplus ({\bf 2} \otimes {\bf 2} \otimes {\bf 2})^\ord{-1} \oplus \scal{ \gl_1 \oplus \sl_2 \oplus \sl_2 \oplus \sl_2 }^\ord{0}  \oplus ({\bf 2} \otimes {\bf 2} \otimes {\bf 2})^\ord{1} \oplus {\bf 1}^\ord{2} \,.
\ee
The corresponding decomposition of the subalgebra and the coset are respectively 
\be
 \bigoplus_{\Lambda= 0}^3 \sl_2^\ord{\Lambda} \cong {\bf 1}^\ord{-2} \oplus \scal{ \gl_1 \oplus \sl_2 \oplus \sl_2 \oplus \sl_2 }^\ord{0} \oplus {\bf 1}^\ord{2} 
\ee
and
\be
 \so(4,4) \ominus \bigoplus_{\Lambda= 0}^3 \sl_2^\ord{\Lambda} \cong  {\bf 2}_0 \otimes {\bf 2}_1  \otimes {\bf 2}_2  \otimes {\bf 2}_3 \cong \scal{ {\bf 2}_1  \otimes {\bf 2}_2  \otimes {\bf 2}_3}^\ord{-1}\oplus \scal{ {\bf 2}_1  \otimes {\bf 2}_2  \otimes {\bf 2}_3}^\ord{1} \,.
\ee
The associated solvable subalgebra is the positive grade component 
\be
 \mathfrak{n}_{\scriptscriptstyle BPS} \cong \scal{ {\bf 2}_1  \otimes {\bf 2}_2  \otimes {\bf 2}_3}^\ord{1} \oplus {\bf 1}^\ord{2} \; , 
\ee
with generators ${\bf e}^\ord{1}_{\alpha_1\alpha_2\alpha_3}\in \mathfrak{p}$ and ${\bf e}^\ord{2}\in\mathfrak{k}^*$ satisfying 
\be
  [ {\bf e}^\ord{1}_{\alpha_1\alpha_2\alpha_3} , {\bf e}^\ord{1}_{\beta_1\beta_2\beta_3} ] = \varepsilon_{\alpha_1\beta_1} \varepsilon_{\alpha_2\beta_2} \varepsilon_{\alpha_3\beta_3} {\bf e}^\ord{2} \,,
\ee
all other commutators being trivial. A generic element is nilpotent of order three in all three fundamental representations, and it follows trivially from (\ref{GradedSystem}) that for the Ansatz 
\be
 \V = \exp\bigl(  L^{\alpha_1\alpha_2\alpha_3}{\bf e}^\ord{1}_{\alpha_1\alpha_2\alpha_3}\bigr)  \,,
\ee
the equations of motion reduces to the free Laplace equation on $L^{\alpha_1\alpha_2\alpha_3}$ 
\be
 d \star d L^{\alpha_1\alpha_2\alpha_3} = 0 \; , 
\ee
and the general solution is determined by 8 harmonic functions on $\mathds{R}^3$. So 
\be
 L^{\alpha_1\alpha_2\alpha_3} = l^{\alpha_1\alpha_2\alpha_3} + \sum_A \frac{q_{\scriptscriptstyle A}^{\alpha_1\alpha_2\alpha_3}}{|x-x_{\scriptscriptstyle A}|} \,.
\ee
At each centre, the existence of a regular horizon requires that the associated Noether charge 
\be
 \QN_{\scriptscriptstyle A} = U_{\scriptscriptstyle A}{}^{-1} q_{\scriptscriptstyle A}^{\alpha_1\alpha_2\alpha_3}{\bf e}^\ord{1}_{\alpha_1\alpha_2\alpha_3}  U_{\scriptscriptstyle A} \; , 
\ee
with 
\be
  U_{\scriptscriptstyle A} =  \exp\left(   \biggl( l^{\alpha_1\alpha_2\alpha_3} + \sum_{B\ne A} \frac{q_{\scriptscriptstyle B}^{\alpha_1\alpha_2\alpha_3}}{|x_{\scriptscriptstyle A}-x_{\scriptscriptstyle B}|}\biggr)  {\bf e}^\ord{1}_{\alpha_1\alpha_2\alpha_3}  \right) \; , 
\ee
lies in the physical orbit, and 
each $q_{\scriptscriptstyle A}$ must for instance admit a positive quartic invariant $I_4>0$ ( defined in \eqref{quarticinvariant} ), that can also be written as
\be
 I_4(q_{\scriptscriptstyle A}^{\alpha_1\alpha_2\alpha_3}) = \varepsilon_{\alpha_1\beta_1}  \varepsilon_{\gamma_1\delta_1}  \varepsilon_{\alpha_2\beta_2}  \varepsilon_{\gamma_2\delta_2}  \varepsilon_{\alpha_3\beta_3}  \varepsilon_{\gamma_3\delta_3}  q_{\scriptscriptstyle A}^{\alpha_1\alpha_2\alpha_3} q_{\scriptscriptstyle A}^{\beta_1\beta_2\beta_3} q_{\scriptscriptstyle A}^{\gamma_1\gamma_2\gamma_3} q_{\scriptscriptstyle A}^{\delta_1\delta_2\delta_3} > 0 
\ee
such that near the horizon $x\approx x_{\scriptscriptstyle A}$ 
\be
 P \in \frac{ SL(2) \times SL(2)\times SL(2)\times SL(2) }{ ( SO(2) \times SO(2) ) \ltimes \mathds{R}^\ord{2}  } \subset \frac{SO(4,4)}{(SO(2) \times SO(2)) \ltimes ( \mathds{R}^{8\, \ord{1}} \oplus  \mathds{R}^{\ord{2}} )} \,.
\ee
However, it is important to note that the total charge 
\be
 \QN = \sum_{A} \QN_{\scriptscriptstyle A} =  \exp(- l^{\beta_1\beta_2\beta_3} {\bf e}^\ord{1}_{\beta_1\beta_2\beta_3}  ) \left( \sum_A q_{\scriptscriptstyle A}^{\alpha_1\alpha_2\alpha_3}{\bf e}^\ord{1}_{\alpha_1\alpha_2\alpha_3}  \right)  \exp( l^{\gamma_1\gamma_2\gamma_3} {\bf e}^\ord{1}_{\gamma_1\gamma_2\gamma_3}  )  
\ee
or equivalently $q = \sum_{A} q_{\scriptscriptstyle A}$, does not necessarily satisfy such constraint, and one can have for example a negative quartic invariant
\be
 \varepsilon_{\alpha_1\beta_1}  \varepsilon_{\gamma_1\delta_1}  \varepsilon_{\alpha_2\beta_2}  \varepsilon_{\gamma_2\delta_2}  \varepsilon_{\alpha_3\beta_3}  \varepsilon_{\gamma_3\delta_3}  q^{\alpha_1\alpha_2\alpha_3} q^{\beta_1\beta_2\beta_3} q^{\gamma_1\gamma_2\gamma_3} q^{\delta_1\delta_2\delta_3} < 0 
\ee
in which case one has
\be
 P \in \frac{ SL(2) \times SL(2)\times SL(2)\times SL(2) }{ ( SO(1,1) \times SO(1,1) ) \ltimes \mathds{R}^\ord{2}  } \subset \frac{SO(4,4)}{(SO(1,1) \times SO(1,1)) \ltimes ( \mathds{R}^{8\, \ord{1}} \oplus  \mathds{R}^{\ord{2}} )} 
\ee
 in the asymptotic region $|x| \rightarrow \infty$, which does not correspond to the asymptotic of any regular single centre black hole. The existence of such solution has been exhibited in \cite{DenefMoore}.

\bigskip
 
The dual vector 
\be
 d W = -  \star d L^{\alpha_1\alpha_2\alpha_3} {\bf e}^\ord{1}_{\alpha_1\alpha_2\alpha_3} +  \varepsilon_{\alpha_1\beta_1} \varepsilon_{\alpha_2\beta_2} \varepsilon_{\alpha_3\beta_3}   L^{\alpha_1\alpha_2\alpha_3} \star d L^{\beta_1\beta_2\beta_3} {\bf e}^\ord{2} 
\ee
includes all the magnetic fields and the Kaluza--Klein vector $\omega$, although the latter is not trivially the component of $W$ along ${\bf e}^\ord{2}$, but a combination which depends on the explicit choice of $h\in \sl_2^\ord{0}$. 

The BPS solutions are well-known and have already been treated in the context of nilpotent orbits in \cite{BossardBPS}, so we will not come back to them.

\subsubsection{The principal orbit algebra: almost-BPS equations} \label{gpmaxorbit}

The maximal nilpotent orbits of $SO_\asym(4,4)$ are associated to $b^\Lambda = (4,2,2,2)$  and its permutations, which means $h=4H_0 +2 H_1+2H_2 +2H_3$, \ie to the weighted Dynkin diagram \DSOVIII2222, which defines the following graded decomposition of $\so(4,4)$
\begin{multline}
 \so(4,4) \cong {\bf 1}^\ord{-5} \oplus {\bf 1}^\ord{-4}  \oplus (3 \times {\bf 1})^\ord{-3} \oplus (3 \times {\bf 1})^\ord{-2}  \oplus (3 \times {\bf 1} \oplus {\bf 1})^\ord{-1} \oplus \\* \scal{ \gl_1 \oplus \gl_1 \oplus \gl_1 \oplus \gl_1 }^\ord{0}\oplus ({\bf 1} \oplus 3 \times {\bf 1} )^\ord{1}    \oplus (3 \times {\bf 1})^\ord{2} \oplus (3 \times {\bf 1})^\ord{3} \oplus {\bf 1}^\ord{4}  \oplus {\bf 1}^\ord{5}  \,.
\end{multline}
For a nilpotent element to be really in the maximal orbit and not in a lower one, it should correspond to a generic element of the grade 2 component, for which the four elements of grade 1 are all non-vanishing. The algebra $\mathfrak{k}^*$ decomposes as
\be
 \sl_2 \oplus \sl_2 \oplus \sl_2 \oplus \sl_2 \cong {\bf 1}^\ord{-4} \oplus  (3 \times {\bf 1})^\ord{-2} \oplus  \scal{ \gl_1 \oplus \gl_1 \oplus \gl_1 \oplus \gl_1 }^\ord{0} \oplus (3 \times {\bf 1})^\ord{2} \oplus {\bf 1}^\ord{4} 
\ee
and accordingly for the coset component
\be
  {\bf 2}_0 \otimes {\bf 2}_1  \otimes {\bf 2}_2  \otimes {\bf 2}_3 \cong {\bf 1}^\ord{-5}  \oplus (3 \times {\bf 1})^\ord{-3}\oplus (3 \times {\bf 1} \oplus {\bf 1})^\ord{-1} \oplus ({\bf 1} \oplus 3 \times {\bf 1} )^\ord{1}    \oplus (3 \times {\bf 1})^\ord{3}   \oplus {\bf 1}^\ord{5} \,. 
\ee
Such element is then nilpotent of order seven in all the three fundamental representations $R_i$ (vector, chiral and antichiral spinor), \ie $\QN_i^{\; 7}=0$, but $\QN_i^{\; 6} \ne 0$.

The relevant solvable algebra is therefore 
\be
 \mathfrak{n}_{\scriptscriptstyle \rm aBPS} \cong ({\bf 1} \oplus 3 \times {\bf 1} )^\ord{1}    \oplus (3 \times {\bf 1})^\ord{2} \oplus (3 \times {\bf 1})^\ord{3} \oplus {\bf 1}^\ord{4}  \oplus {\bf 1}^\ord{5} \,.
\ee
We will label ${\bf e}^\ord{1}_i$ and ${\bf e}^\ord{1}_0$ the grade one generators,   ${\bf e}^\ord{2}_i$ the grade two, ${\bf e}^{\ord{3} i}$ the grade three, ${\bf e}^\ord{4}$ and  ${\bf e}^\ord{5}$ the grade four and five. The non-vanishing commutator can be written in terms of the symmetric $\IR_+^* \times \IR_+^*$ invariant tensor $c_{ijk} \equiv | \varepsilon_{ijk}|$, as 
\bea
 && [ {\bf e}^\ord{1}_0 , {\bf e}^\ord{1}_i ] = {\bf e}^\ord{2}_i \qquad  [ {\bf e}^\ord{2}_i , {\bf e}^\ord{1}_j ] =  c_{ijk} {\bf e}^{\ord{3} k} \qquad  [ {\bf e}^{\ord{3} i} ,   {\bf e}^\ord{1}_j ] = \delta^i_j {\bf e}^\ord{4} \CR 
 && [ {\bf e}^\ord{4} , {\bf e}^\ord{1}_0 ] = {\bf e}^\ord{5} \qquad   [ {\bf e}^\ord{2}_i ,  {\bf e}^{\ord{3} j} ] = \delta_i^j {\bf e}^\ord{5} \,. \label{comutmax} 
\eea
Only the generators of odd grade are in the coset component, so one can consider the Ansatz 
\be
 \V = \exp\scal{ - \tilde V {\bf e}^\ord{1}_0 - \tilde K^i {\bf e}^\ord{1}_i  - \tilde Z_i {\bf e}^{\ord{3} i} - \tilde M {\bf e}^\ord{5} } \,.
\ee
After some algebra, one obtains the odd
\begin{multline}
  - P = d \tilde V {\bf e}^\ord{1}_0 + d \tilde K^i {\bf e}^\ord{1}_i + \left( d \tilde Z_i + \frac{1}{6} c_{ijk} \scal{ \tilde K^j \tilde K^k d\tilde  V - \tilde V \tilde K^j d\tilde  K^k } \right)  {\bf e}^{\ord{3} i} \\* + \left( d \tilde M + \frac{1}{6} \scal{ \tilde V \tilde K^i d\tilde  Z^\prime_i - 2 \tilde V\tilde Z_i d \tilde K^i + \tilde  Z_i \tilde K^i d\tilde V } + \frac{1}{120} c_{ijk} \tilde V \tilde K^j \tilde K^k \scal{ \tilde K^i d \tilde V -\tilde  V d \tilde K^i } \right)  {\bf e}^\ord{5} 
\end{multline}
and even component of the Maurer--Cartan form $\V^{-1} d \V= P + B  $
\be
 B = \frac{1}{2} \scal{  \tilde K^i d\tilde  V- \tilde V d \tilde K^i  } {\bf e}^\ord{2}_i + \frac{1}{2} \left(   \tilde K^i d \tilde Z_i -\tilde  Z_i d \tilde K^i + \frac{1}{12} c_{ijk} \tilde K^j \tilde K^k \scal{  \tilde K^i d \tilde V-  \tilde V d \tilde K^i} \right)     {\bf e}^\ord{4} \,.
\ee
 
The grade one component of the equations of motion (\ref{eom2}) gives obviously that $\tilde V$ and $\tilde K^i$ are harmonic functions:
\be
 d \star d V = d \star d K^i =0 \,.  
\ee
Using the harmonicity of $\tilde V$ and $\tilde K^i$, one reduces the grade three component to 
\be \label{eqalBPSZ}
 d \star d \left( \tilde Z_i + \frac{1}{6} c_{ijk} \tilde V \tilde K^j \tilde K^k \right)  = \frac{1}{2} c_{ijk} \tilde V d \star d  \scal{ \tilde K^j \tilde K^k } 
\ee
and using all these equations combined, the grade five component gives 
\be \label{eqalBPSM}
 d \star d  \left( \tilde M + \frac{2}{3} \tilde V \tilde Z_i \tilde K^i + \frac{1}{15} \tilde V^2 c_{ijk} \tilde K^i \tilde K^j \tilde K^k  \right) = 2 d \left(\tilde  V \Scal{  \tilde Z_i + \frac{1}{6} c_{ijk}\tilde  V \tilde K^j \tilde K^k} \star d\tilde  K^i \right) \,.
\ee
It follows that with the redefinitions
\bea \label{redefVZiM}
 a V &\equiv&  \tilde V + l_0 \,, \qquad K^i \equiv \tilde K^i \,, \CR
a Z_i &\equiv&  \left(  \tilde Z_i + \frac{1}{6} c_{ijk} ( \tilde  V + 3 l_0 ) \tilde K^j \tilde K^k \right) + l_i \,, \\
a^2 V \mu &\equiv& \frac{1}{2} \tilde  M + \biggl( \frac{1}{3} + l_0 \tilde Z_i + \tilde V l_i - l_0 l_i \biggr)  \tilde V \tilde Z_i \tilde K^i  + \frac{1}{6} \biggl( \frac{1}{5}  \tilde{V}{}^2 + \frac{1}{2} l_0  ( \tilde V - l_0) \biggr)  c_{ijk} \tilde K^i\tilde  K^j \tilde K^k \,, \nonumber 
\eea
one obtains the almost-BPS equations, \cite{Goldstein:2008fq,Bena:2009ev}, for arbitrary constant $a$ and $l_\Lambda$. It should be recalled that this system of equations was initially found in a completely different way, and it seems quite remarkable that we recover it with our approach. However, one of the physical assumptions behind this system being extremality, it is natural that it admits a description in terms of nilpotent orbits, which we exhibit here. This provides a group theoretical explanation for the graded structure of the almost-BPS system.
\medskip

A generic element  $q^0  {\bf e}^\ord{1}_0 + q^i {\bf e}^\ord{1}_i $ of the grade one component does not commute with any generator of $\bigoplus_\Lambda \sl_2^\ord{\Lambda}$, and therefore defines a representative of a maximal nilpotent orbit. It satisfies $\QN_i{}^7=0$. An element with $q^0 = 0 $, $ q^i {\bf e}^\ord{1}_i + p_i  {\bf e}^{\ord{3}\, i} + p_0 {\bf e}^\ord{5}$ lies in the BPS nilpotent orbit $\QN_i{}^3 = 0 $ $b^\Lambda = (2,0,0,0)$, and can be chosen such that it corresponds to a regular single-centre BPS black hole. With one or two vanishing $q^i$, the charge generically lies in a nilpotent orbit $\QN_i{}^5 = 0$ and $\QN_j{}^4 =0$ for $i\ne j$. With only one non-vanishing $q^i$, the other components can be chosen such that all $\QN_i{}^3 =0$, but one can then check that the corresponding configuration never corresponds to a regular black hole because the required reality conditions are not satisfied.\footnote{By this we mean that the Levi subgroup of the stabilizer of $\QN_i$ in $K^*$ is not compact.} For $q^i=0$, the element is generically nilpotent  $\QN_i{}^3 = 0 $ with $b^\Lambda = (1,1,1,1)$, and can be chosen such that it corresponds to a regular single-centre black hole. These requirements can be understood in terms of  type IIA bound states in a particular duality frame in which $q^0$ can be identified to a \Db{6} charge, $q^i$ to D4 charges, $p_i$ to D2 charges and $p_0$ to a D0 charge.

\subsubsection{The subregular orbit algebra:  composite non-BPS equations} \label{gpsubmaxorbit}

The next to maximal nilpotent orbits of $SO_\asym(4,4)$ are associated to $b^\Lambda = (0,2,2,2)$, $h= 2H_1+2H_2 +2H_3$ and its permutations. Its weighted Dynkin diagram is \DSOVIII2220 and it defines the following graded decomposition of $\so(4,4)$
\be
 \so(4,4) \cong {\bf 2}^\ord{-3} \oplus (3 \times {\bf 1})^\ord{-2} \oplus (3 \times {\bf 2})^\ord{-1} \oplus \scal{ \gl_1 \oplus \gl_1 \oplus \gl_1 \oplus \sl_2 }^\ord{0} \oplus (3 \times {\bf 2})^\ord{1}\oplus (3 \times {\bf 1})^\ord{2} \oplus {\bf 2}^\ord{3} \,.
\ee
The nilpotent elements correspond to linearly independent doublet of 3-vector of non-vanishing cubic invariant, of either equal sign or not. The algebra $\mathfrak{k}^*$ decomposes as
\be
 \sl_2 \oplus \sl_2 \oplus \sl_2 \oplus \sl_2 \cong  (3 \times {\bf 1})^\ord{-2} \oplus  \scal{ \gl_1 \oplus \gl_1 \oplus \gl_1 \oplus \sl_2 }^\ord{0} \oplus (3 \times {\bf 1})^\ord{2} 
\ee
and accordingly
\be
  {\bf 2}_0 \otimes {\bf 2}_1  \otimes {\bf 2}_2  \otimes {\bf 2}_3 \cong  {\bf 2}^\ord{-3}\oplus (3 \times {\bf 2})^\ord{-1} \oplus (3 \times {\bf 2})^\ord{1} \oplus {\bf 2}^\ord{3} \,.
\ee
Such element satisfies $\QN_i^{\; 5} = 0$, but $\QN_i^{\; 4} \ne 0$.  

The relevant solvable subalgebra is 
\be
 \mathfrak{n}_{\scriptscriptstyle \rm nBPS} \cong  (3 \times {\bf 2})^\ord{1}\oplus (3 \times {\bf 1})^\ord{2} \oplus {\bf 2}^\ord{3} \,.
\ee
We will label the grade one generators ${\bf e}^i_\alpha$, the grade two generators ${\bf e}_k$ and the grade three generators ${\bf e}_\alpha$. They satisfy the algebra
\be
 [ {\bf e}^i_\alpha , {\bf e}^j_\beta ] = \varepsilon_{\alpha\beta} c^{ijk} {\bf e}_k \qquad [ {\bf e}_i ,  {\bf e}^j_\alpha ] = \delta_i^j {\bf e}_\alpha \,,
\ee
with $c^{ijk}=|\varepsilon^{ijk}|$. The relevant Ansatz is then 
\be
 \V = \exp\scal{ - K_i^\alpha  {\bf e}^i_\alpha - M^\alpha {\bf e}_\alpha } \,.
\ee
One computes that 
\be
 - P = d K_i^\alpha  {\bf e}^i_\alpha + \left( d M^\alpha  - \frac{1}{6} c^{ijk} K_i^\alpha \varepsilon_{\beta\gamma} K^\beta_j d K^\gamma_k \right) {\bf e}_\alpha 
\ee
and 
\be
 B = -\frac{1}{2} \varepsilon_{\alpha\beta} c^{ijk} K_i^\alpha d K_j^\beta {\bf e}_k 
\ee
and the equations of motion imply that $K^i_\alpha$ are all harmonic 
\be
 d \star d K_i^\alpha =0
\ee 
and 
\be
 d \star d M^\alpha = \frac{2}{3} d \Scal{ \varepsilon_{\beta\gamma} c^{ijk} K_i^\alpha K_j^\beta \star d K_k^\gamma } \,.
\ee 

This gives us a {\it new system of equations}, that we call the composite non-BPS system, and we will show in section \ref{compositenonBPSsection} and \ref{solvecompositenonBPS} that it admits new non-BPS multi-centre interacting solutions. 

To finish this section, we note that, as for the almost-BPS case, one can recover the physical non-BPS orbit $b^\Lambda = (1,1,1,1)$ as a subcase: a charge $\QN = q_i^\alpha {\bf e}^i_\alpha + p^\alpha {\bf e}_\alpha $ corresponds to a physical non-BPS orbit, verifying $\QN_i{}^3 = 0 $, if 
\be
 \varepsilon_{\alpha\beta} q_i^\alpha q_j^\beta  = 0   
\ee
and 
\be
  \forall i \, \quad \varepsilon_{\alpha\beta} p^\alpha q_i^\beta > 0 \,.
\ee
It is straightforward to check that one can chose several charges $\QN_\pA$ satisfying these requirement such that $\QN= \sum \QN_\pA$ defines a generic element of the nilpotent orbit associated to $b^\Lambda= (0,2,2,2)$. The algebra analysis therefore already suggests that interacting solutions exist.

\section{From the group theory to the physical system} \label{sec3}

In this section we will present all the necessary material required for defining explicit four-dimensional solutions from the algebraic solutions we have described in the preceding section. We will also discuss the expression of the ADM mass in function of the asymptotic central charges. 

\subsection{Conventions}

Our aim is to describe black hole solutions of the STU model. Since this model has already been studied in great details we will only briefly recall some properties of the theory, but refer the reader for example to \cite{Behrndt:1996hu,BossardW,Tdualpaper} for further details. The STU model is a particularly useful truncation of 11-dimensional supergravity, as well as the type II supergravity theories.  We will sometimes refer to the microstates interpretation of black hole solutions in terms of D-branes in type IIA supergravity. The model can be understood as $\N=1$ supergravity in five dimensions coupled to three vector multiplets with the intersection form $c_{ijk} = |\varepsilon_{ijk}|$. More details about the uplift of the BPS and almost-BPS systems can be found in \cite{Tdualpaper}, whereas the uplift to five dimensions of the new composite non-BPS system will be presented in section \ref{compositenonBPSsection}. 

\medskip

We will consider the standard Ansatz for the stationary metric
\be \label{4Dmet}
 ds^2 = - e^{2U} \bigl( dt + \omega \bigr)^2 + e^{-2U}\delta_{\mu\nu} dx^\mu dx^\nu  \; . 
\ee
The 8 electromagnetic fields (4 electric $A^\Lambda$ plus 4  magnetic duals $A_\Lambda$) of the STU model decompose accordingly as
\be \label{4DEM}
-  2 \sqrt{2} \, A_{A_1B_2C_3} = \zeta_{A_1B_2C_3}  \bigl( dt + \omega \bigr) +  w_{A_1B_2C_3} \; , 
\ee
where each $A_i, B_i, C_i$ runs from $1$ to $2$ of the corresponding $SL(2,\IR)_i$ duality symmetry associated to the complex scalars
\be \label{4Dscal}
 t^i = a_i + i e^{-2\phi_i} \,, \qquad i=1,2,3\; ,
\ee
parametrizing the upper complex half plan $U(1)_i \backslash  SL(2,\IR)_i$. As we have already discussed, the stationary equations of motion reduce to the ones of Euclidean 3-dimensional gravity coupled to a non-linear sigma model defined on $G/K^* \cong SO(4,4)/ ( SO(2,2) \times SO(2,2))$, \ie \eqref{eom1}, \eqref{eom2}: 
\be
 \trace P_\mu P_\nu = R_{\mu\nu} \quad  , \qquad  
 d \star P +  [ B , \star P ] = 0 \; , 
\ee
where, as we saw in the previous section, $P$ and $B$ are respectively the coset and subgroup component of the Maurer-Cartan one-form $\V^{-1} d\V$ of the coset representative $\V \in SO(4,4)/ ( SO(2,2) \times SO(2,2))$. The analysis of these equations has been done in the previous section. Here, we focus on how the four-dimensional fields \eqref{4Dmet}, \eqref{4DEM}, \eqref{4Dscal} are encoded inside the three-dimensional coset element $\V$.


Using the standard convention that capital indices $A_i, B_i, \dots $ correspond to rigid $SL(2)_i$ (acting on the left) whereas small ones $a_i, b_i, \dots$ to local $SO(2)$ (acting on the right), we define $SO(4,4)$ as the subgroup of $SL(8,\IR)$ preserving the metric
\be \label{defeta}
 \upeta = \left(\begin{array}{ccc} 
\hspace{2mm} 0  \hspace{2mm}& \hspace{2mm} \varepsilon_{A_1B_1}  \hspace{2mm}  & \hspace{2mm}0 \hspace{2mm}  \\
\hspace{2mm} - \varepsilon_{A_1B_1}   \hspace{2mm}& \hspace{2mm}0  \hspace{2mm}  & \hspace{2mm} 0  \hspace{2mm}  \\
\hspace{2mm}  0    \hspace{2mm}& \hspace{2mm}  0 \hspace{2mm}  & \hspace{2mm} \varepsilon^{A_2B_2} \varepsilon^{A_3B_3} \hspace{2mm} \end{array} \right)  \begin{array}{c} \\ \\ \vspace{4mm} 
\end{array} \; , 
\ee
and the coset representative  $\V \in SO(4,4)/(SO(2,2) \times SO(2,2))$ as
\bea \label{defV}
   \V &=&  \exp\bigl[ \zeta^{A_1B_2C_3} {\bf E}_{A_1B_2C_3} + \sigma {\bf E} \bigr] \exp[ U {\bf H}] 
\left(\begin{array}{ccc} 
\hspace{2mm} v_{a_1}{}^{A^\prime_1}  \hspace{2mm}& \hspace{2mm} 0  \hspace{2mm}  & \hspace{2mm}0 \hspace{2mm}  \\
\hspace{2mm}0  \hspace{2mm}& \hspace{2mm} v_{a_1}{}^{A_1^\prime}  \hspace{2mm}  & \hspace{2mm} 0  \hspace{2mm}  \\
\hspace{2mm}  0    \hspace{2mm}& \hspace{2mm}  0 \hspace{2mm}  & \hspace{2mm} (v^{-1})_{B_2^\prime}{}^{b_2} (v^{-1})_{C_3^\prime}{}^{c_3} \hspace{2mm}
 \end{array} \right)  \begin{array}{c} \\ \\ \vspace{4mm} \end{array} \\*
 &=& \left(\begin{array}{ccc} 
\hspace{2mm} e^U v_{a_1}{}^{A_1}  \hspace{2mm}& \hspace{2mm} e^{-U} v_{a_1}{}^{A_1} \sigma + \frac{1}{2} e^{-U} v_{a_1}{}^{B_1}{}  \zeta^{A_1D_2E_3} \zeta_{B_1D_2E_3}     \hspace{2mm}  & \hspace{2mm} (v^{-1})_{D_2}{}^{b_2} (v^{-1})_{E_3}{}^{c_3}  \zeta^{A_1D_2E_3} \hspace{2mm}  \\
\hspace{2mm}0  \hspace{2mm}& \hspace{2mm}e^{-U} v_{a_1}{}^{A_1}  \hspace{2mm}  & \hspace{2mm} 0  \hspace{2mm}  \\
\hspace{2mm}  0    \hspace{2mm}& \hspace{2mm}  e^{-U} v_{a_1}{}^{D_1}  \zeta_{D_1B_2C_3} \hspace{2mm}  & \hspace{2mm} (v^{-1})_{B_2}{}^{b_2} (v^{-1})_{C_3}{}^{c_3} \hspace{2mm}
 \end{array} \right) \nonumber 
\eea
which must be understood as acting on a vector $( A^{a_1} , B^{a_1} , C_{b_2 c_2})$. Note also that the order of the indices for $v_{a_1}{}^{A_1}$ is reversed because it is the transverse of $v_{a_1}{}^{A_1}$ which is involved. A completely explicit form of this conventions is given in the Appendix. 

\medskip

The $v_{a_i}{}^{A_i}$ are the $SO(2) \backslash SL(2,\IR)$ representatives 
\be
(v_{a_i}{}^{A_i}) \equiv \left(\begin{array}{cc} 
\; v_1{}^1\; & \; v_1{}^2 \; \\ \, v_2{}^1  \, & \;  v_2{}^2  \; 
\end{array}\right)   = \left(\begin{array}{cc} 
\; e^{-\phi_i} \; & \; 0 \; \\ \, e^{\phi_i} a_i \, & \;  e^{\phi_i} \; 
\end{array}\right) \, ;
\ee
$ \zeta^{A_1B_2C_3}$ define the electromagnetic fields, and indices are raised and lowered via $\varepsilon_{A_iB_i}$, \eg
\be \label{zetahtbas}
 \zeta_{A_1B_2C_3} = \varepsilon_{A_1D_1}  \varepsilon_{B_2E_2}  \varepsilon_{C_3F_3}  \zeta^{D_1E_2F_3} \; . 
\ee 
We parametrize the electromagnetic fields in terms of 2 by 4 matrices
\be \label{def2by4EM}
 ( \zeta^{A_1B_2C_3} )  \equiv \left(\begin{array}{cccc} 
\, \zeta^{111}\,  & \, \zeta^{112} \, & \; \zeta^{121} \; & \; \zeta^{122} \; \\
\; \zeta^{211} \; & \; \zeta^{212} \; & \, \zeta^{221} \, & \, \zeta^{222} \, 
\end{array}\right) = 
 \left(\begin{array}{cccc} 
\, \psi_0\,  & \, -\psi_3 \, & \; -\psi_2\; & \; -\chi^1 \; \\
\; -\psi_1 \; & \; -\chi^2 \; & \, -\chi^3 \, & \, -\chi^0 \, 
\end{array}\right) \; ,   
\ee
and in terms of 4 by 2 matrices from \eqref{zetahtbas},
\be \label{def4by2EM}
 (\zeta_{A_1B_2C_3})  \equiv \left(\begin{array}{cc} 
\, \zeta_{111} \,  & \; \zeta_{211} \; \\
\; \zeta_{112} \; & \; \zeta_{212} \; \\ \, \zeta_{121} \, & \, \zeta_{221} \, \\ \; \zeta_{122} \; & \; \zeta_{222} \;   
\end{array}\right)  
= \left(\begin{array}{cc} 
\, -\chi^0\,  & \; \chi^1 \; \\
\; \chi^3 \; & \; -\psi_2 \; \\ \, \chi^2 \, & \, -\psi_3 \, \\ \; -\psi_1 \; & \; -\psi_0 \;   
\end{array}\right) \; . 
\ee
Triality is realised as the permutation of the three indices $i=1,2,3$ in this basis.

\medskip

The involution defining the $K^*$ subgroup is defined from the twisted transpose  
\be
 \V^\ddagger = \left(\begin{array}{ccc} 
\hspace{2mm} \delta_{a_1}^{a^\prime_1}  \hspace{2mm}& \hspace{2mm}0 \hspace{2mm}  & \hspace{2mm}0 \hspace{2mm}  \\
\hspace{2mm} 0   \hspace{2mm}& \hspace{2mm} \delta_{a_1}^{a^\prime_1}  \hspace{2mm}  & \hspace{2mm} 0  \hspace{2mm}  \\
\hspace{2mm}  0    \hspace{2mm}& \hspace{2mm}  0 \hspace{2mm}  & \hspace{2mm}  - \delta_{b_2^\prime}^{b_2} \delta_{c_3^\prime}^{c_3}  \hspace{2mm} \end{array} \right)  \begin{array}{c} \\ \\ \vspace{4mm} \end{array} \V^{\rm T}  \left(\begin{array}{ccc} 
\hspace{2mm} \delta^{A_1}_{A^\prime_1}  \hspace{2mm}& \hspace{2mm}0 \hspace{2mm}  & \hspace{2mm}0 \hspace{2mm}  \\
\hspace{2mm} 0   \hspace{2mm}& \hspace{2mm} \delta^{A_1}_{A^\prime_1}  \hspace{2mm}  & \hspace{2mm} 0  \hspace{2mm}  \\
\hspace{2mm}  0    \hspace{2mm}& \hspace{2mm}  0 \hspace{2mm}  & \hspace{2mm}  - \delta^{B_2^\prime}_{B_2} \delta^{C_3^\prime}_{C_3}  \hspace{2mm} \end{array} \right)  \begin{array}{c} \\ \\ \vspace{4mm} \end{array} \; , 
\ee
such that $\V^\ddagger = \V^{-1}$ iff $\V \in K^*$. The relation to the representative  $\exp(-L)$ in the symmetric gauge \eqref{VeqexpL} discussed in the previous section is 
\be \label{VVdagmat}
 \exp(-2L) = \V  \V^\ddagger =  \left(\begin{array}{ccc} 
\hspace{2mm} \times  \hspace{2mm}& \hspace{2mm} e^{-2U} M^{A_1B_1} \sigma + \frac{1}{2} e^{-2U} M^{C_1B_1}{}  \zeta^{A_1D_2E_3} \zeta_{C_1D_2E_3}     \hspace{2mm}  & \hspace{2mm} \times \hspace{2mm}  \\
\hspace{2mm} \times   \hspace{2mm}& \hspace{2mm}e^{-2U} M^{A_1B_1}  \hspace{2mm}  & \hspace{2mm} \times   \hspace{2mm}  \\
\hspace{2mm}  \times     \hspace{2mm}& \hspace{2mm}  e^{-2U} M^{A_1D_1}  \zeta_{D_1B_2C_3} \hspace{2mm}  & \hspace{2mm} \times  \hspace{2mm} \end{array} \right) \; , 
\ee
where 
\be
 M^{A_iB_i}  = \left(\begin{array}{cc} e^{-2\phi_i} + e^{2\phi_i} a_i^{\; 2}  & \;  e^{2\phi_i} a_i  \; \\ \; e^{2\phi_i} a_i \; & \;  e^{2\phi_i} \; 
\end{array}\right) \; , 
\ee
and we only wrote the elements of the middle column because they are enough to determine all the fields, when using triality. We recall that the interest of looking at $\V  \V^\ddagger$ instead of $\V$ is that $\V \V^\ddagger$ is, by construction, invariant under $[SL(2,\IR)]^4$ gauge transformations. From \eqref{VVdagmat} it is then easy to decode the three-dimensional matrix to obtain the four-dimensional scalars. To obtain the full four-dimensional fields, one finally needs to dualize some of them to obtain the expressions of the vectors. 

From the equations of motion, the Kaluza--Klein vector $\omega$ is defined by
\be \label{eqomega}
 d \omega = \trace {\bf E} \, dW  = \star  e^{-4U} \biggl( d \sigma - \frac{1}{2} \zeta^{A_1B_2C_3} d \zeta_{A_1B_2C_3} \biggr)  \; , 
\ee
and the magnetic vectors by
\bea \label{eqemvectors}
 d w_{A_1B_2C_3} \!&=& - \frac{1}{4} \trace {\bf E}_{A_1B_2C_3} d W  \CR
&=& e^{-2U} \! M^{-1}_{A_1D_1} M^{-1}_{B_2E_2} M^{-1}_{C_2F_2} \star  d \zeta^{D_1E_2F_2} \! - \! e^{-4U} \! \star \! \biggl( d \sigma \! - \! \frac{1}{2} \zeta^{D_1E_2F_2} d \zeta_{D_1E_2F_2} \biggr)  \zeta_{A_1B_1C_1} \CR
&=& e^{-2U}  M^{-1}_{A_1D_1} M^{-1}_{B_2E_2} M^{-1}_{C_2F_2} \star  d \zeta^{D_1E_2F_2}  - \zeta_{A_1B_2C_3} d \omega \; , 
\eea
Note that the electromagnetic field strength are then manifestly twisted self-dual
\be
 - 2 \sqrt{2} \, F_{A_1B_2C_3} = d \zeta_{A_1B_2C_3} {}_{\wedge}  \bigl( dt + \omega \bigr) + e^{-2U}  M^{-1}_{A_1D_1} M^{-1}_{B_2E_2} M^{-1}_{C_2F_2} \star  d \zeta^{D_1E_2F_2} \; . 
\ee
This ends the parametrization of the physical fields in terms of the three-dimensional coset element $\V$.

 In order to obtain solutions associated to a given solvable system as discussed in the previous sections, we must now provide an explicit basis of function $L$ lying in corresponding solvable algebra   $\mathfrak{n} \cap \mathfrak{p}$.  In this aim, we define a particular basis of Cartan generators of the subalgebra $\bigoplus_\Lambda \sl_2^\ord{\Lambda}$ (according to the matrix notation) as 
\be \label{defHLambda}
 H_\Lambda \equiv \left( \begin{array}{cc} 
\; 0 \; & \; U_\Lambda \; \\ \; U_\Lambda^{\; {\rm T}} \; & \; 0 \; 
\end{array}\right) \; , 
\ee
where $U_\Lambda$ are the four specific 4 by 4 matrices 
\be
 U_\Lambda \equiv  \left(\begin{array}{cccc} 
\, \delta_\Lambda^0 - \frac{1}{2}  \, & \; 0 \; & \; 0\; & \, \delta_\Lambda^1 - \frac{1}{2} \, \\
\; 0 \; & \, \delta_\Lambda^2 - \frac{1}{2} \, &  \, \delta_\Lambda^3 - \frac{1}{2} \, & \; 0 \; \\
\; 0 \; & \, \frac{1}{2} - \delta_\Lambda^3 \, &  \,  \frac{1}{2} - \delta_\Lambda^2  \, & \; 0 \; \\
\,  \frac{1}{2} - \delta_\Lambda^1  \, & \; 0 \; & \; 0\; & \, \frac{1}{2} - \delta_\Lambda^0 \, 
\end{array} \right) \label{OffDiagH} \; .
\ee
%
%
The generic elements $X$ of $\mathfrak{p}$ are defined similarly as (\ref{CentralCharge}) below. One can easily solve 
\be
 [ H_\Lambda , X_\beta ] =  \beta_\Lambda X_\beta \; , \label{Heigen} 
\ee
for $\beta_\Lambda = ( \pm 1, \pm1,\pm1,\pm1)$, which defines a basis for the 16 elements of $\mathfrak{p}$.  The explicit form of the elements $X_\beta$ is given in the appendix. The functions $L$ of \eqref{VeqexpL} in the positive grade components of $\mathfrak{p}$ associated to a solvable system are then simply obtained in this basis as
\begin{itemize}
\item  $\displaystyle L = \sum_{\beta_0 = 1} L^\beta X_\beta$ in the BPS system;
\item  $\displaystyle L = \sum_{2\beta_0 + \sum_i \beta_i \, \ge\,  1} L^\beta X_\beta$ in the almost-BPS system;
\item  $\displaystyle L = \sum_{\sum_i \beta_i \, \ge \, 1} L^\beta X_\beta$ in the composite non-BPS system;
\end{itemize}
respectively. 
 
As explained in section \ref{GSS}, the most general solutions of these types are generated by the action of $\KK \cong [U(1)]^4$. The general Cartan basis $H_\Lambda$ can be obtained by conjugating (\ref{OffDiagH}) by a general $[U(1)]^4$ rotation, given explicitly in the appendix, such that  $H_\Lambda(\alpha_\Lambda)$ is a function of the phase $\alpha_\Lambda$ only, by property of the $\bigoplus_\Lambda \sl_2^\ord{\Lambda}$ algebra. Nevertheless it will be easier to first compute the explicit basis $X_\beta$ for specific $H_\Lambda$ and then rotate the associated function $L=\sum L^\beta X_\beta$ with respect to $\KK$, rather than to compute the explicit basis for the general $H_\Lambda(\alpha_\Lambda)$ from scratch.


\subsection{Charges and central charges} \label{chargessec}

We are now going to describe how the central charges of the solution can easily be extracted from the coset representative $\V$, and how the mass formula can be determined from the algebraic conditions satisfied by the solutions. This will permit to generalise the standard BPS formula $M=|Z|$ to the non-BPS systems.  

\medskip

Considering any space-like cycle $\Sigma$  embedded in $M_4$ through $\iota$, orthogonal to the time-like vector $\partial_t$, the associated electromagnetic charge is
\be
 q_{A_1B_2C_3\, |\Sigma} = \! \frac{1}{2\pi} \int_{\iota (\Sigma)} \!\!\!\! \iota^*  F_{A_1B_2C_3}  = \! - \frac{1}{4\pi \sqrt{2}} \int_{\Sigma} \!\! d \scal{ w_{A_1B_2C_3} + \omega \zeta_{A_1B_2C_3} } = \!  \frac{1}{4 \sqrt{2}} \trace {\bf E}_{A_1B_2C_3}  \QN_{|\Sigma} \; 
\ee
where we used the absence of Dirac--Misner string 
\be
 \trace {\bf E}  \, \QN_{|\Sigma} = 0 \; , 
\ee
to show that the contribution from $\zeta_{A_1B_2C_3}\,\omega$ drops out.

 It is straightforward to compute that the total Noether charge $\QN$ can be rotated to the coset component using the asymptotic value of the scalar fields $\V_\asym$ 
 \be 
\V_\asym{}^{-1} \QN \V_\asym =  \left(\begin{array}{ccc} 
\hspace{2mm} M  \delta^{a_1}_{d_1} + \sigma_{d_1}{}^{a_1}  \hspace{2mm} & \hspace{2mm} N  \delta^{a_1}_{d_1}  \hspace{2mm}  & \hspace{2mm} z^{a_1e_2f_3} \hspace{2mm}  \\
\hspace{2mm}  N  \delta^{a_1}_{d_1}   \hspace{2mm} & \hspace{2mm} - M \delta^{a_1}_{d_1} + \sigma_{d_1}{}^{a_1}  \hspace{2mm}  & \hspace{2mm} - z_{a_1e_2f_3}   \hspace{2mm}  \\
\hspace{2mm}  -z^{d_1b_2c_3}    \hspace{2mm} & \hspace{2mm}  z_{d_1b_2c_3}  \hspace{2mm}  & \hspace{2mm}- \sigma_{b_2}{}^{e_2} \delta_{c_3}^{f_3} - \delta_{b_2}^{e_2}  \sigma_{c_3}{}^{f_3} \hspace{2mm}
 \end{array} \right) , \label{CentralCharge} 
\ee
defined as a matrix acting on a vector $(A^{d_1}, B^{d_1}, C_{e_2f_2})$. Here, the (2,3) component $(-z_{a_1e_2f_3})$ and (3,1) $(-z^{d_1b_2c_3})$ can be understood as matrices, as minus the transpose of $(z^{a_1e_2f_3})$ and $(z_{d_1b_2c_3})$ defined as in \eqref{def2by4EM},\eqref{def4by2EM}. $M$ is the ADM mass of space-time, the total NUT charge $N$ is assumed to vanish, $\sigma_{a_i}{}^{b_i}$ are symmetric traceless matrices 
\be
(\sigma_{a_i}{}^{b_i}) \equiv\left(\begin{array}{cc} 
\; \Sigma_i \; & \; \Xi_i \; \\ \, \Xi_i  \, & \,- \Sigma_i \, 
\end{array}\right) \, ,
\ee
associated to the momenta of the scalar fields in the asymptotic region
\be 
\Pi_i \equiv - \lim_{r\rightarrow \infty}  \frac{r^2 \partial_r  t^i }{t^i- \bar t^i} = - \Sigma_i + i \Xi_i  \; ,  
\ee
and
\be
 z^{a_1b_2c_3} \equiv  v_\asym^{a_1}{}_{A_1} v_\asym^{b_2}{}_{A_2} v_\asym^{c_3}{}_{C_3} q^{A_1B_2C_3} 
\ee 
defines the $[SL(2,\IR)]^3$ covariant combination of the asymptotic central charge and its asymptotic K\"{a}hler derivatives, which we will call the asymptotic `central charges'. The asymptotic `central charges' are defined in terms of $z^{a_1b_2c_3}$ as
\be
 Z = \frac{1}{2} \sum_{a_1,a_2,a_3} (-i)^{a_1+a_2 +a_3} z^{a_1a_2a_3} \; , \qquad Z_{i} = \frac{1}{2}   \sum_{a_1,a_2,a_3} (i)^{a_i} (-i)^{a_{i+1} +a_{i+2} } z^{a_1a_2a_3}\; . 
\ee
We have 
\be
 Z = \frac{1}{\sqrt{ i \prod_j ( t^j_\asym - \bar t^j_\asym)} } \Bigl( q_0 + \sum_i t^i_\asym q_i + \sum_i t^{i+1}_\asym t^{i+2}_\asym p^i - t^1_\asym t^2_\asym t^3_\asym p^0 \Bigr) \; , 
\ee
and the $Z_i$'s are obtained by replacing $t^i_\asym$ by its complex conjugate in the holomorphic component. 

\medskip

It will be illuminating to discuss the action of $\KK \cong [U(1)]^4$ on the asymptotic momenta we just defined. These rotations act on $\V_\asym{}^{-1} \QN \V_\asym$ given in (\ref{CentralCharge}) such that 
\be
\begin{split}  \tilde{M}-i\tilde{N} &= e^{\frac{i}{2} ( \alpha_0 + \alpha_1 + \alpha_2 + \alpha_3 ) } ( M - i N ) \\
\tilde \Pi_i & = e^{\frac{i}{2}( - \alpha_0 - \alpha_i + \alpha_{i+1}  + \alpha_{i+2} ) }  \Pi_i 
\end{split}\qquad
\begin{split}
\tilde Z & = e^{\frac{i}{2} ( - \alpha_0 + \alpha_1 + \alpha_2 + \alpha_3 ) } Z \\
 \tilde Z_i & = e^{\frac{i}{2}( - \alpha_0 + \alpha_i - \alpha_{i+1}  - \alpha_{i+2} ) }  Z_i 
\end{split} \label{AMRotation} 
\ee
Considering a solution associated to a given nilpotent orbit, $ \V_\asym{}^{-1} \QN \V_\asym$ lies by definition in the closure of this nilpotent orbit. It follows that it lies in the positive grade component defined by a generator $h  = b^\Lambda H_\Lambda(\alpha_\Lambda)$ for some phases $\alpha_\Lambda$. We now have to specify the $b^\Lambda$ \ie the orbit we are in, to obtain an explicit relation. One can already anticipate that the number of phases it depends on is simply the number of nonzero $b^\Lambda$ labelling the orbit. 

For the BPS case, it is well known that one has~\footnote{Here we shift all phases $\alpha_\Lambda$ by $\frac{\pi}{2}$ with respect to the convention (\ref{OffDiagH}) to avoid the presence of extra $i$ factors.}
\be
 M - i N = e^{-i \alpha_0} Z \; , \qquad  \Pi_i = e^{i \alpha_0} \bar Z_i \; . 
\ee
This translates the fact that only $b^0 \neq 0$ in the BPS orbit. Requiring the absence of NUT charge determines the phase $\alpha_0$ as the one of the asymptotic central charge such that $M = |Z|$.

We now turn to the almost-BPS orbit: in order to obtain the general form of the asymptotic momenta in function of the asymptotic central charges for the non-BPS systems, it is convenient to first compute the constraint at $\alpha_\Lambda = 0 $, and then obtain the general solution by applying the transformations (\ref{AMRotation}). The property that the almost-BPS system can be obtained from the BPS system by replacing a D6 charge by \Db6, illustrates into the fact that the ADM mass and asymptotic scalar momenta are given by 
\be 
M - i N =  Z - \frac{1}{2} \mbox{Re} \Bigl[ Z + \sum_i \bar Z_i \Bigr]   \; , \qquad  \Pi_i =  \bar Z_i  - \frac{1}{2} \mbox{Re} \Bigl[ Z + \sum_j \bar Z_j \Bigr]\; . 
\ee
With respect to the BPS case, the new term comes from the presence of the \Db6 instead of the D6. One can indeed check, using \eqref{D0D2D4D6}, that for a pure \Db6 charge, one obtains $M= - Z$. Reintroducing the angles, we see that in this case the ADM mass is not determined by the asymptotic `central charges' only, but depends on three phases $\alpha_i$ such that 
\be
 M = \frac{1}{4} \Bigl( 3 e^{-i\alpha_0} Z - e^{-i \sum_i \alpha_i } \bar Z - \sum_i \bigl( e^{-i ( \alpha_0 + \alpha_{i+1} + \alpha_{i+2} ) } Z_ i + e^{-i\alpha_i } \bar Z_i \bigr) \Bigr) \; , 
\ee
where $\alpha_0$ is determined in function of the $\alpha_i$'s such that this expression is real positive, and in particular such that the NUT charge vanishes. 

The composite non-BPS system asymptotic momenta can themselves be obtained starting from the almost-BPS system by substituting \Db4 charges to the D4 charges as
\bea
 M - i N &=& Z - \frac{1}{2}  \mbox{Re} \Bigl[ Z + \sum_i \bar Z_i \Bigr]  - \frac{i}{2} \sum_i \mbox{Im} \Bigl[ Z - \bar Z_i +\sum_{j\ne i} \bar Z_{j} \Bigr] \CR
&=& \frac{1}{2} \Bigl( \bar Z - \sum_i \bar Z_i \Bigr) \; ,  
\eea
and 
\bea
 \Pi_i &=& \bar Z_i - \frac{1}{2}  \mbox{Re} \Bigl[ Z + \sum_i \bar Z_i \Bigr]  + \frac{i}{2}  \mbox{Im} \Bigl[ Z - \bar Z_i +\sum_{j\ne i} \bar Z_{j} \Bigr] -  \frac{i}{2} \sum_{j\ne i} \mbox{Im} \Bigl[ Z - \bar Z_j + \sum_{k\ne j} \bar Z_{k} \Bigr]  \CR
&=& \frac{1}{2} \Bigl( - Z + Z_i - \sum_{j\ne i} \bar Z_{j} \Bigr) \; . 
\eea
It is remarkable that although the expression of the ADM mass and asymptotic scalar momenta are rather complicated deformations of the BPS expressions, they end up being pretty simple. Restoring the dependency in the phases one gets 
\bea \label{nonBPSADM}
 M &=&  \frac{1}{2} \Bigl( e^{-i \sum_i \alpha_i} \bar Z - \sum_i e^{-i\alpha_i} \bar Z_i \Bigr) \; , \\
 \Pi_i &=&  \frac{1}{2} \Bigl( - e^{i\alpha_i} Z +e^{i(\alpha_i - \alpha_{i+1} - \alpha_{i+2})}  Z_i - e^{-i\alpha_{i+2}} Z_{i+1} - e^{-i\alpha_{i+3}}  Z_{i+2} \Bigr) \; ,  
\eea
where the phases $\alpha_i$ are again assumed to be chosen such that the ADM mass is real positive. Note that the dependence in $\alpha_0$ drops out, which comes from the fact that $b^0=0$ in the composite non-BPS case, and thus the associated rotation leaves invariant $h = 2 \sum_i H_i(\alpha_i)$. It is remarkable that the ADM mass reduces to the ADM mass formula computed in \cite{Gimon:2007mh} within a specific Ansatz up to a four-dimensional duality transformation parametrized by the three phases $\alpha_i$. It follows that the ADM mass of a  bound state of non-BPS black holes is defined by the generalised `fake superpotential' defined in \cite{Ceresole:2009vp}, for specific values of the `auxiliary fields' associated to the flat directions which will depend in general on the specific interior structure of the solution. Note that the `auxiliary fields' are extremum of the generalised `fake superpotential' if and only if there is no interactions between the centres. 

To finish, let us discuss the case of single-centre non-BPS black holes, that are particular solutions of both the almost-BPS and the composite non-BPS system. They furthermore satisfy that 
\be
 \mbox{Im}\bigl[ e^{\frac{i}{2}( - \alpha_0 + \alpha_i - \alpha_{i+1} - \alpha_{i+2})} Z_i \bigr] = \mbox{Im}\bigl[ e^{\frac{i}{2}( - \alpha_0 + \alpha_1+  \alpha_{2} + \alpha_{3})} Z \bigr]   \; ,  
\ee
which corresponds to the absence of D4 or \Db4 charges respectively. All the phases $\alpha_\Lambda$'s are then determined in function of the asymptotic `central charges' as in the BPS case, in accordance with the no-hair theorem. One computes straightforwardly that the duality rotation 
\be
 \tilde Z = e^{i \sum_i ( \alpha_i - \frac{1}{4} \sum_\Lambda \alpha_\Lambda + \frac{3\pi}{4})} Z \; , \qquad \tilde  Z_i =  e^{i ( \alpha_i - \frac{1}{4} \sum_\Lambda \alpha_\Lambda + \frac{3\pi}{4})} Z_i \; , 
\ee
provides a solution to the non-standard diagonalization problem which defines the  `fake superpotential' \cite{BossardW} (with $\alpha = \frac{1}{4} \sum_\Lambda \alpha_\Lambda$ in (2.68) of \cite{BossardW}), such that (\ref{nonBPSADM}) indeed reduces to the asymptotic value of the `fake superpotential'. The equation of the asymptotic scalar momenta implies the existence of two flat directions
\be
 \mbox{Im}\bigl[ e^{\frac{i}{2}( - \alpha_0 - \alpha_i + \alpha_{i+1} + \alpha_{i+2})} \Pi_i \bigr]  = - \mbox{Im}\bigl[ e^{\frac{i}{2}( - \alpha_0 + \alpha_1+  \alpha_{2} + \alpha_{3})} Z \bigr]   \; . 
\ee

\medskip

It is important to understand that the action of $\KK \cong [U(1)]^4$ on the solvable systems themselves is very similar to its action on the asymptotic momenta. From the stabilizer of the defining semi-simple element $h \equiv b^\Lambda H_\Lambda$ one obtains similarly that the BPS system is parametrized by one single phase $\alpha_0$, the almost-BPS system by the four phases $\alpha_\Lambda$, and the composite non-BPS system by the three phases $\alpha_i$. The discussion of the last paragraph suggests that there is always one combination of the parametrizing phases which does not permit to obtain new interesting systems, as far as asymptotically Minkowski composite black hole solutions are concerned. Indeed, the BPS system is known to be unique. We will see in the following that the Ehlers rotation of the almost-BPS system can also be reabsorbed in a reparametrization of the solution, excepted for a particular value of the phase which gives rise to a system which does not admit regular composite black hole solutions. It follows that the general almost-BPS system is in fact parametrized by three phases associated to the compact subgroup of the four-dimensional duality group. The Ehlers $U(1)$ acts already trivially on the composite non-BPS system, and there is no further restriction in this case, such that the three phases then parametrize inequivalent classes of solutions.

\section{Almost BPS solutions}

We have seen in section \ref{gpmaxorbit} that the equations associated to the maximal orbit of $\so(4,4)$ are exactly the almost-BPS equations \cite{Goldstein:2008fq,Bena:2009ev}. Here, using the tools from the previous section, we will relate the system of equations, obtained from a purely algebraic point of view, to physical solutions of the STU model. We will first show how to recover all the known almost-BPS solutions \cite{Bena:2009ev,Bena:2009en,Bena:2009fi}, including their ``generalised version" \cite{Tdualpaper,Bena:2011zq}, and show that the Ehlers rotation does not produce any new solutions than the one obtained by the four-dimensional dualities.

\subsection{Recovering the almost-BPS usual solutions}

We now want to decode the four-dimensional fields out of $\mathcal{M}=\mathcal{V}\eta\mathcal{V}^t\eta$. This is of course not a one-to-one correspondence, because the fields from the three-dimensional perspective are duality invariants. One must thus choose a duality frame in order to write an explicit solution. It is interesting to choose the one where $V$ correspond to a \Db{6}-charge and $K^i$ to D4-charges. Indeed, this is the duality frame where the first almost-BPS solutions have been found, and is therefore well-known \cite{Bena:2009ev}. This corresponds naturally to the parametrization given in the previous section and in the appendix \eqref{defHLambda}, \eqref{D0D2D4D6}. The general solution should include all the dependences in the $\KK$ angles, but, for clarity, we will not consider it here. It should however be clear that the three free phases of $\KK$ are sufficient to obtain the whole class of almost-BPS solutions discovered in \cite{Tdualpaper}, and this in a straightforward way. 

In order to recover the almost-BPS solution in its original parametrization, it is interesting to use the freedom of redefinition of \eqref{redefVZiM}. We here assume for convenience the constants in \eqref{redefVZiM} to be $a=-2$ and $l_0=l_i=-2$:
\bea
  V &\equiv&  -\frac{1}{2}\left( \tilde V -2 \right) \qquad K^i \equiv \tilde K^i \CR
  Z_i &\equiv&  -\frac{1}{2} \left(  \tilde Z_i + \frac{1}{6} c_{ijk} ( \tilde  V - 6 ) \tilde K^j \tilde K^k \right) -2 \\
  V \mu &\equiv& \frac{1}{4} \left( \frac{1}{2} \tilde M + ( 2 - \ti{V} )  \ti{K}^i + \Bigl( -1 + \frac{\ti{V}}{3} \Bigr) \ti{Z}_i \ti{K}^i +  \Bigl( \frac{1}{5} \tilde{V}^2 - 2 \tilde V + 4 \Bigr)  \tilde K^1 \tilde  K^2 \tilde K^3 \right) \,. \nonumber 
\eea 
With these choices, After having chosen this particular duality frame, the four-dimensional scalar fields can be extracted from $\V$ using \eqref{VVdagmat}:
\bea \label{alBPS4Dfields}
 \rm{e}^{-4U} &=& V Z_1 Z_2 Z_3 - V^2 \mu^2 \,, \nonumber \\
 \rm{e}^{-2 \phi_i} &=& \frac{\rm{e}^{-2U}}{V Z_i} \,, \nonumber \\
 a_i &=& K^i - \frac{\mu}{Z_i} \,, \nonumber \\
 \chi^0 &=&  \rm{e}^{4U} V^2 \mu \,, \\ \nonumber
 \chi^i &=& 1 + \rm{e}^{4U}V \bigl(  - Z_{i+1} Z_{i+2} + V K^i \mu \bigr) \,, \\ \nonumber
 \psi_0 &=& -1 + \rm{e}^{4U}\left(  Z_1 Z_2 Z_3 - V\mu \sum_i Z_i K^i + V \sum_{i<j} Z_i Z_j K^i K^j \right) \,, \\ \nonumber
 \psi_i &=& \rm{e}^{4U} V \left( \big( Z_i + V K^{i+1} K^{i+2} \big) \mu - Z_i\sum_{j\neq i}K^j Z_j \right) \,, \\ \nonumber
 \sigma &=& \frac{ \rm{e}^{4U} V }{2} \left( (-2+V)\mu +\mu \sum_i \big( Z_i +  V K^{i+1} K^{i+2} \big) - \sum_{i\neq j} K^i Z_i Z_j\right) \,.
\eea
This describes exactly the solutions of \cite{Bena:2009ev,Bena:2009en}.\footnote{Because these fields do not strictly speaking define tensors, we have not used Einstein summation convention in these equations. We recall that these solutions have been first written in a five-dimensional language. The map to its four-dimensional version is done in \cite{Tdualpaper}.}

\paragraph{Vectors.}

For completeness, one also have to check that the vectors are the same. These ones are given by the equations \eqref{eqomega}, \eqref{eqemvectors}. Having chosen a particular duality frame, it will be more convenient to write the electro magnetic fields as two four-vectors. The map from the $2 \times 2 \times 2$ to two four-vectors is given by \eqref{def2by4EM}, with the equivalent for the vectors $w^\Lambda$, $v_\Lambda$. \footnote{Note that we do not introduce the $-2\sqrt{2}$ factor in the definition of the four-dimensional vectors as in (\ref{4DEM}) in order to facilitate the comparison with \cite{Bena:2009ev,Bena:2009en}.}
\bea \label{defALambda}
 A^\Lambda &=& \chi^\Lambda ( dt + \omega ) + w^\Lambda \,, \\ \nonumber
 A_\Lambda &=& \psi_\Lambda ( dt + \omega ) + v_\Lambda \,,
\eea
The $\omega$ equation rewrites
\bea \label{galeqomega2}
 \star d \omega = \rm{e}^{4U} \left( d \sigma -\frac{1}{2} \big( \chi^\Lambda d \psi_\Lambda - \psi_\Lambda d \chi^\Lambda \big) \right) \,.
\eea
Plugging in the values for our system \eqref{alBPS4Dfields}, this equation greatly simplifies, to give
\bea
 \star d \omega &=& d(V \mu) - V Z_i d K^i \,.
\eea
%

%
%
%
One can also see that using \eqref{alBPS4Dfields}, the equations for the electromagnetic vectors \eqref{eqemvectors} simplify drastically:
\bea
 \star d w^0 &=& dV \,, \qquad \star dv_0 = Z_i dK^i - K^i dZ_i +\left( V d(K^1 K^2 K^3) - K^1 K^2 K^3 dV \right) \,, \\ \nonumber
 \star dw^i &=& -\star d \omega -V d K^i + K^i dV \,, \qquad \star dv_i = dZ_i + \frac{c_{ijk}}{2} \left( K^j K^k dV - V d(K^j K^k) \right) \,.
\eea
Up to a small gauge choice, this is exactly the equations for the vectors of \cite{Tdualpaper}. Therefore one completely recovers the known almost-BPS solutions.

\subsection{Ehlers rotation}

In \cite{Tdualpaper}, the authors have used the four-dimensional duality group $SL(2,\IR)^3$ of the STU model to produce new solutions. As we have already seen, the transformation that can be done with this group are part of $\KK$. However, in three dimensions, there is another $U(1)$, coming from the Ehlers reduction along time. In our framework, it is thus clear that one can reobtain all solutions of \cite{Tdualpaper}, by acting with $\KK$ on $\V \V^\ddagger$. In principle one could also obtain new solutions with the extra Ehlers $U(1)$ action, but we will see that this is not the case, as advocated in the previous section. 

To show this, we perform a rotation with angles\footnote{The additional rotations along the $SL(2,\IR)_i, i=1,2,3$ are done by convenience, and do not change the argumentation. The precise parametrization of the rotation is given in the appendix.} $\alpha_0=\alpha_1=\alpha_2=\alpha_3=\alpha/2$
\be
 (\V \V^\ddagger)' = R(\alpha) \V \V^\ddagger R(\alpha)^{\; {\rm T}} \,.    
\ee
The expressions for the physical fields are then {\it a priori} a complicated mixture of all involved functions, and $\alpha$ factors. However, one can do the following redefinitions
\bea
 V &=&  \frac{1}{a}\left( \tilde V +l_0 \right) \,, \qquad K^i = \frac{1}{c}\tilde K^i + \sin\alpha \,,\CR
  Z_i &=&  \frac{1}{a \, c^2} \left(  \tilde Z_i + \frac{1}{3}  ( \tilde  V + 3 l_0 ) \tilde K^{i+1} \tilde K^{i+2} \right) + l_i + \sum_{j\neq i} 2 \ti{K}^j \tan\alpha \,, \\
  V \mu &=& \frac{1}{a^2 c^3} \left( \frac{1}{2}( l_0 + \ti{V} ) \sum_i l_i \ti{K}^i + \Bigl( \frac{l_0}{2} + \frac{\ti{V}}{3} \Bigr) \sum_i \ti{Z}_i \ti{K}^i +  \Bigl( \frac{1}{5} \tilde{V}^2 +l_0 \tilde V + l_0^2 \Bigr)   \tilde K^1 \tilde  K^2 \tilde K^3 \right.  \nonumber \\ \nonumber
 && \left. \hspace{-10mm}- 2 \tan\alpha \Bigl( l_0 + \frac{2\ti{V}}{3} \Bigr) \sum_{i<j} \ti{K}^i \ti{K}^j - \tan \alpha \sum_i \ti{Z}_i -\tan\alpha \, \ti{V} + 2 \tan^2\!\alpha \sum_i \ti{K}^i + m_0 +  \frac{1}{2} \tilde M \right) \,.
\eea
These redefinitions seem to be rather complicated, but the point is that they are just a generalisation of \eqref{redefVZiM}, where we had an extra harmonic function in the $Z_i$'s and in $M$, and an additive and a multiplicative factor to the $K_i$'s, but one can easily check that this is still compatible with the almost-BPS system of equations \eqref{eqalBPSZ},\eqref{eqalBPSM}, with the constants being given by
\be
 a  = -\frac{2}{\cos^2\!\alpha} \,, \qquad c = \cos \alpha \,, \quad l_0 = l_i = -\frac{2}{\cos\alpha} \,, \qquad m_0 = 4\frac{\sin\alpha}{\cos^2\!\alpha} \,.
\ee
These redefinitions degenerate for $\alpha=\pi/2$, but one can check in this case that there are no physical solutions. Then e$^{-4U}$ and the scalar fields are exactly given by \eqref{alBPS4Dfields} :
\bea
 \rm{e}^{-4U} = V Z_1 Z_2 Z_3 - V^2 \mu^2 \,, \qquad  \rm{e}^{-2 \phi_i} = \frac{\rm{e}^{-2U}}{V Z_i} \,, \qquad a_i = K^i - \frac{\mu}{Z_i} \,.
\eea
The electromagnetic fields are also, up to a constant, the same as in \eqref{alBPS4Dfields}
\bea 
 \chi^0 &=& -\sin\alpha + \rm{e}^{4U} V^2 \mu  \,, \\ \nonumber
 \chi^i &=&  \cos\alpha + \rm{e}^{4U}V\bigl(  -Z_{i+1} Z_{i+2} + V K^i \mu \bigr)  \,, \\ \nonumber
 \psi_0 &=& -\frac{1 + \sin^2\!\alpha}{\cos\alpha} + \rm{e}^{4U}\biggl( Z_1 Z_2 Z_3 - V\mu \sum_i  Z_i K^i + V \sum_{i<j} Z_i Z_j K^i K^j \biggr)  \,, \\ \nonumber
 \psi_i &=&  \sin\alpha + \rm{e}^{4U} V \left( \big( Z_i +V K^{i+1} K^{i+2} \big) \mu - Z_i\sum_{j\neq i}K^j Z_j \right) \,.
\eea
So we are in the same class of solutions as before. Note that, because of the different constants in the electromagnetic fields, the explicit expression of $\sigma$  differs from \eqref{alBPS4Dfields}, but they coincide up to a gauge transformation, and are therefore physically equivalent. We conclude that the Ehlers rotation does not produce any new solutions, but led us in the class already spanned by the four-dimensional duality group.

\section{Composite non-BPS solutions}  \label{compositenonBPSsection}

We now turn to the study of the composite non-BPS class of solutions unveiled in section \ref{gpsubmaxorbit}. Because this class was not known so far, we present it in more details. We will focus on the four-dimensional solution, although the uplift to five dimensions will also be shortly discussed. In the next section, we will exhibit an explicit two-centre solution in this class, and study its physical properties.

\subsection{Equations of motion}

This system represents solutions where half of the branes have been inverted. The major difference with the almost-BPS system is that our system is now ``symmetric'' in terms of the two triplets of functions ${K_i}^{\alpha=1}$ and ${K_i}^{\alpha=2}$, whereas the latter was not in terms of $K^i$ and $Z_i$. It is however convenient to break this manifest symmetry by redefining the following functions 
\bea \label{redeffctsubmax}
 \ti{K}_i^{\alpha=1} &=& 2(L_i +1) \,, \nonumber \\
 \ti{K}_i^{\alpha=2} &=& K_i \,, \\ \nonumber
 \ti{M}^{\alpha=1} &=& 8 M + \frac{4}{3} \left( \sum_i K_i + \sum_{i\neq j} K_i L_j - 3 \sum_i K_i L_{i+1} L_{i+2} \right) \,, \\ \nonumber
 \ti{M}^{\alpha=2} &=& 2 (V+1) + \frac{1}{3} \left( 2 \sum_{i \neq j} K_i K_j -3 \sum_i K_{i+1} K_{i+2} L_i \right) \,.  
\eea
With these redefinitions, the system of equations becomes
\bea \label{eqscalars}
 d\star d \, K_i &=& 0 \,, \nonumber \\
 d\star d \, L_i &=& 0 \,,  \\ \nonumber
 d\star d \, M &=& \frac{1}{2} c^{ijk}  d \bigl(  L_i L_j \star d K_k \bigr) \,, \\ \nonumber
 d\star d \, V &=& \frac{1}{2} c^{ijk} d \bigl(  L_i \star d (K_j K_k) - K_j K_k \star d L_i \bigr)  \,.
\eea
%

\subsection{Reconstructing the solution}

\subsubsection{Four dimensional point of view}

As we have done for the almost-BPS solutions in the previous section, we want here to reconstruct the four-dimensional solution from the $\V \V^\ddagger$ matrix. We will not present the most general solution, but restrict, for the sake of clarity, to one particular duality frame. In order to compare easily with the results of the almost-BPS class of solutions, we uplift the solution in the duality frame where ($V,K_i,L_i,M$) correspond respectively to the (\Db{6},\Db{4},D2,D0)-charges. In particular, for $K_i=0$, one falls back into the physical non-BPS orbit, describing single-centre \Db{6}-D2-D2-D2 black holes. In our duality frame, the redefinitions \eqref{redeffctsubmax} are such that e$^{-4U}$ takes the simple usual expression 
\bea \label{em4Usm}
 \rm{e}^{-4U} = V L_1 L_2 L_3 - M^2 \,.
\eea
Introducing the quantities
\bea \label{defTi}
 T_i \equiv V L_i - 2 K_i M  + K_i^2 L_{i+1} L_{i+2}  \,, 
\eea
the scalar fields are given by
\bea \label{saclsm}
t^i = \frac{ K_iL_{i+1}L_{i+2}   - M + i \rm{e}^{-2U}  }{ T_i }  \,.
\eea
The electromagnetic fields and $\sigma$ are finally given by
\bea \label{EMsigmasm}
 \chi^0 &=& \rm{e}^{4U}\Bigl( -M V + \sum_i  \big( V L_{i+1} L_{i+2} K_i - M K_{i+1} K_{i+2} L_i \big) + L_1 L_2 L_3 K_1 K_2 K_3 \Bigr) \,, \nonumber \\
 \chi^i &=& -1 + \rm{e}^{4U} \Bigl( M ( K_{i+1}  L_{i+2}+ K_{i+2}  L_{i+1})   - L_{i+1} L_{i+2} ( V+ L_i K_{i+1} K_{i+2} ) \Bigr) \,, \nonumber \\
 \psi_0 &=& -1 - \rm{e}^{4U} L_1 L_2 L_3 \,,  \\ \nonumber
 \psi_i &=& -\rm{e}^{4U}L_i\bigl( -M +  K_i L_{i+1} L_{i+2} \bigr)  \,, \\ \nonumber
 \sigma &=& \frac{ \rm{e}^{4U} }{2} \Big( -M \Big( 2 + V + \sum_i L_i \Big) + \sum_i \big( V L_{i+1} L_{i+2} K_i - M K_{i+1} K_{i+2} L_i \big) \\ \nonumber
 && + L_1 L_2 L_3 \sum_i K_i + L_1 L_2 L_3 K_1 K_2 K_3 \Big) \,. 
\eea

\paragraph{Vectors.}

Having chosen a duality frame, one can compute the explicit equations for the vectors. The general equations are given in \eqref{eqomega},\eqref{eqemvectors}. Using the explicit values for our system \eqref{em4Usm}-\eqref{EMsigmasm}, these equations greatly simplify, to give
\bea \label{eqomegasm}
 \star d \omega = d M - \frac{1}{2}c^{ijk} L_i L_j d K_k 
\eea
for $\omega$, and
\bea \label{EMvectorssm}
 \star d w^0 &=& - d V + \frac{1}{2} c^{ijk} \big( L_i d (K_j K_k) - K_j K_k d L_i \big) \,, \nonumber \\
 \star d w^i &=& \star d \omega + c^{ijk} \big( K_j d L_k - L_k d K_j \big) \,, \\ \nonumber
 \star d v_0 &=& \star d \omega =  d M - \frac{1}{2} c^{ijk} L_i L_j d K_k \,, \\ \nonumber
 \star d v_i &=& d L_i \,,
\eea
for the electromagnetic vectors. One can check, as it should be, that the compatibility equations are nothing but the scalar equations \eqref{eqscalars}.

\subsubsection{Uplift to five dimensions}

It is interesting to uplift the solution to five dimensions. In this framework, the fields are simply the five dimensional metric $ds_5^2$, the three electromagnetic one-forms $A^i_5$ and three real scalars $X^i$, $i=1,2,3$ verifying $X^1X^2X^3=1$. They are related to the four-dimensional fields through
\bea \label{uplift}
 ds_5^2 &=& \rm{e}^{4 \Phi /3} ( d\psi - A^0 )^2 + \rm{e}^{-2\Phi /3} ds_4^2 \,, \nonumber \\
 A^i_5 &=& A^i + \rm{Re}(t^i) ( d\psi - A^0 ) \,, \\ \nonumber
 X^i &=& \rm{e}^{-2\Phi /3}\rm{Im}(t^i) \,,
\eea
where $\psi$ is the fifth coordinate, and we defined $\Phi$ by
\be
 \rm{e}^{2\Phi} \equiv \rm{Im}(t^1)\rm{Im}(t^2)\rm{Im}(t^3) \,.
\ee
Applied explicitly to the solution of the last subsection, this gives the five-dimensional metric
\be \label{5Dmet}
 ds_5^2 = - \frac{H}{T^2} \left( dt + \omega - \rm{e}^{-4U} \frac{\chi^0}{H}(d\psi - w^0) \right)^2 + \frac{T}{H}(d\psi - w^0)^2 + T ds_3^2 \,, 
\ee
where $T$ is defined as $T \equiv (T_1 T_2 T_3)^{1/3} $, and $H$ by
\bea
 H \equiv \Bigl( V - \frac{c^{ijk}}{2}K_i K_j L_k\Bigr)^2 + 4 K_1 K_2 K_3 \Bigl(2M - \frac{c^{ijk}}{2} K_i L_j L_k\Bigr) \,. 
\eea
The one-forms are given by
\begin{multline}
 A^i_5 = dt - \frac{1}{ T_i} \Bigl( V + \bigl( L_i K_{i+1} K_{i+2} -  K_i K_{i+1} L_{i+2}- K_i K_{i+2} L_{i+1}  \bigr) \Bigr) \bigl(  dt + \omega \bigr) \\*
 +  w^i - \omega + \frac{ K_i L_{i+1} L_{i+2} -  M}{ T_i} \bigl( d\psi - w^0\bigr)  \,.
\end{multline}
In this duality frame, the presence of the $d\psi - w^0$ term in the metric implies a non-trivial fibration of the associated circle over the four-dimensional base in the presence of a non-vanishing \Db{6} charge. In particular, fixing the \Db6 charge to 1 and choosing appropriate boundary conditions such that $T^2$ and $H$ both scale as $r^{-2}$ in the asymptotic region should permit to define asymptoticaly Minkowki solutions. It should be interesting to investigate this question.

\subsection{Rotation along time}

We show here that, as expected from the fact that $b^0=0$ for the composite non-BPS system, the Ehlers rotation does not produce any new solutions, and thus that the whole class of composite non-BPS solutions is parametrized by the four-dimensional duality group. In fact, the rotation along the time direction uses the symmetry in $\alpha=1,2$. To show this, starting from the general $\V\V^\ddagger$ given in \eqref{VVdagmat}, we perform a rotation, with general parameter $\alpha/2$ along the $U(1)_0$ in $\KK$, together with a rotation along each of the three $U(1)_i$ with parameter $-\alpha/2$   
\bea
 (\mathcal{\V\V^\ddagger})' = R(\alpha) \V\V^\ddagger R(\alpha )^T \,.
\eea
Then, if one redefines
\bea
 \ti{K}^1_i &=& 2 \bigl( (1+L_i) \cos\alpha + K_i \sin\alpha \bigr) \,, \nonumber \\
 \ti{K}^2_i &=& 2 \bigl(  K_i \cos\alpha - (1+L_i) \sin\alpha \bigr) \,, \nonumber \\
 \ti{M}^1 &=& \left( 8 M + \frac{4}{3} \left( \sum_i K_i + \sum_{i \neq j} K_i L_j - 3 \sum_i L_{i+1} L_{i+2} K_i \right) \right) \cos \alpha \\ \nonumber
 && + \left( 4(1+V) + \frac{4}{3} \left( \sum_{i \neq j}K_i K_j -3 \sum_i  L_i K_{i+1} K_{i+2} \right) \right) \sin \alpha \,, \\ \nonumber
 \ti{M}^2 &=& \left( 2(1+V) + \frac{2}{3} \left( \sum_{i \neq j}K_i K_j - 3 \sum_i  L_i K_{i+1} K_{i+2} \right) \right) \cos \alpha \\ \nonumber
 && - \left( 4 M + \frac{2}{3} \left( \sum_i K_i + \sum_{i \neq j} K_i L_j - 6 \sum_i  L_{i+1} L_{i+2} K_i \right) \right) \sin \alpha \,,
\eea
one first can show that the new functions still solve the system of equation \eqref{eqscalars}, and that one recovers exactly the fields given in the previous subsection. This proves that the rotation along the Ehlers $SL(2,\IR)$ just transforms a solution into another of the same class.

\section{Solving the composite non-BPS system} \label{solvecompositenonBPS}

In the previous section, we have written the four-, and five-dimensional Ans\"atze associated to the new system of equations \eqref{eqscalars}, and we will now solve the equations explicitly. As we have explained previously, for a solution to be into this new orbit, and be regular, it has to have more than one centre. Indeed, the single-centre solutions are forced by regularity to be in the BPS or the physical non-BPS orbit. We will not discuss the single-centre non-BPS solution in this paper, and refer to \cite{Gaiotto:2007ag,Bena:2009ev} for an explicit exposition of the latter. For the sake of simplicity, we will only construct here an axisymmetric two-centre solution.\footnote{Note that a two-centre non-BPS solution is not axisymmetric if the intrinsic angular mementa of the black holes are not parallel to the line which joins them.} Similarly as for the almost-BPS class of solutions, we are going to see that the system admits regular solutions that do not belong to any previously known subsystem.  Although regularity implies that the coset momenta $P$ falls back into the physical non-BPS orbit at each centre, it belongs to the higher order subregular nilpotent orbit at a generic point of the three-dimensional base subspace. The centres are then interacting, in the sense that their electromagnetic charges produce an angular momentum and the distance between the centres is fixed.

\subsection{The solution}

One parametrizes the three dimensional base space in spherical coordinates $(r,\theta,\phi)$ and take the two centres to be at $r=0$ for the first one and along the positive $z$ axis ($\theta=0$) at a distance $R$ from the origin for the second one. We denote the polar coordinates centred at the second centre position as $(\Sigma,\theta_\Sigma)$. Their relation to the polar coordinates $(r,\theta)$ centred at the origin is:
\be
\Sigma = \sqrt{r^2 + R^2 - 2 r R \cos\theta}\,,\qquad \cos\theta_\Sigma = {r\cos\theta-R\over \Sigma}\,.
\label{polarSigma}
\ee
Looking at a general two-centre solution, we take the $L_i$'s and $K_i$'s to be of the form
\bea \label{LandK}
 L_i = l_i + \frac{Q_i}{r} + \frac{\ti{Q}_i}{\Sigma} \,, \quad K_i = k_i + \frac{d_i}{r} + \frac{\ti{d}_i}{\Sigma} \,.
\eea
We now have to solve the equations \eqref{eqscalars} for $V$ and $M$. Let's first look at the equation for $M$. From \eqref{LandK}, it can be rewritten as
\bea
 d \star d M = \frac{c^{ijk}}{2} d \left[ \frac{2 l_i Q_j}{r} + \frac{2 l_i \ti{Q}_j}{\Sigma} + \frac{Q_i Q_j}{r^2} + \frac{\ti{Q}_i \ti{Q}_j}{\Sigma^2} + \frac{2 Q_i \ti{Q}_j}{r \Sigma} \right] \wedge \star d \left[ \frac{d_k}{r} + \frac{\ti{d}_k}{\Sigma} \right] \,.
\eea
It will be convenient to decompose these linear equations into the several pieces 
\bea
 d \star d M_1 &=& d \big( \frac{1}{r} \big) \wedge \star d \big( \frac{1}{r} \big) \,, \quad d \star d M_2 = d \big( \frac{1}{r^2} \big) \wedge \star d \big( \frac{1}{r} \big) \,, \quad d \star d M_3 = d \big( \frac{1}{r} \big) \wedge \star d \big( \frac{1}{\Sigma} \big) \,, \nonumber \\ 
 d \star d M_4 &=& d \big( \frac{1}{r^2} \big) \wedge \star d \big( \frac{1}{\Sigma} \big) \,, \quad  d \star d M_5 = d \big( \frac{1}{r \Sigma} \big) \wedge \star d \big( \frac{1}{r} \big) 
\eea
and symmetric expressions in $r\leftrightarrow\Sigma$. By direct computation, the solutions to these equations are
\bea
 M_1 = \frac{1}{2 r^2} \,, \quad M_2 = \frac{1}{3 r^3} \,, \quad M_3 = \frac{1}{2 r \Sigma} \,, \quad M_4 = \frac{\cos \theta}{R r \Sigma} \,, \quad M_5 = -\frac{\cos \theta_\Sigma}{2 R r^2} \,.
\eea
We obtain symmetric expressions in $r$ and $\Sigma$ by exchanging $(r,R,\cos\theta)$ with $(\Sigma,-R,\cos\theta_\Sigma)$.
Finally, including the allowed harmonic contribution for $M$, the solution is
\bea
 M &=& m_0 + \frac{m}{r} + \frac{\ti{m}}{\Sigma} + \alpha \frac{\cos \theta}{r^2} + \ti{\alpha} \frac{\cos \theta_\Sigma}{\Sigma^2} \nonumber \\
 && + \frac{c^{ijk}}{2} \Big[ \, \frac{l_i Q_j d_k}{r^2} + \frac{l_i \ti{Q}_j \ti{d}_k}{\Sigma^2} + \frac{Q_i Q_j d_k}{3 r^3} + \frac{\ti{Q}_i \ti{Q}_j \ti{d}_k}{3 \Sigma^3} + \frac{l_i ( Q_j \ti{d}_k + \ti{Q}_j d_k ) }{r \Sigma}  \\ \nonumber
 && + Q_i Q_j \ti{d}_k \frac{\cos \theta}{R r \Sigma } - \ti{Q}_i \ti{Q}_j d_k \frac{\cos \theta_\Sigma}{R r \Sigma } - Q_i \ti{Q}_j d_k \frac{\cos \theta_\Sigma}{R r^2} + Q_i \ti{Q}_j \ti{d}_k \frac{\cos \theta}{R \Sigma^2 } \, \Big] \,.
\eea
Because one can rewrite the equation for $V$ as 
\be
 d \star d \big( V - \frac{c^{ijk}}{2} K_i K_j L_k \big) = - c^{ijk} d (K_i K_j ) \wedge \star d L_k \,, 
\ee
the structure is exactly the same as for $M$, multiplying the terms by $-2$ and exchanging $K_i$ and $L_i$. Hence, the solution for $V$ is
\bea
 V &=& \frac{c^{ijk}}{2} L_i K_j K_k + h + \frac{Q_6}{r} + \frac{\ti{Q}_6}{\Sigma} + \beta \frac{\cos \theta}{r^2} + \ti{\beta} \frac{\cos \theta_\Sigma}{\Sigma^2} \nonumber \\
 && - c^{ijk} \Big[ \, \frac{k_i d_j Q_k}{r^2} + \frac{k_i \ti{d}_j \ti{Q}_k}{\Sigma^2} + \frac{d_i d_j Q_k}{3 r^3} + \frac{\ti{d}_i \ti{d}_j \ti{Q}_k}{3 \Sigma^3} + \frac{k_i ( d_j \ti{Q}_k + \ti{d}_j Q_k )}{r \Sigma}  \\ \nonumber
 && + d_i d_j \ti{Q}_k \frac{\cos \theta}{R r \Sigma } - \ti{d}_i \ti{d}_j Q_k \frac{\cos \theta_\Sigma}{R r \Sigma } - d_i \ti{d}_j Q_k \frac{\cos \theta_\Sigma}{R r^2} + d_i \ti{d}_j \ti{Q}_k \frac{\cos \theta}{R \Sigma^2 } \, \Big] \,.
\eea
To obtain the full solution, we finally need to solve \eqref{eqomegasm} and \eqref{EMvectorssm} for the vectors $\omega$, $w^\Lambda$ and $v_\Lambda$. Clearly, the terms $M_1$ and $M_2$ in $M$ involving only one centre will give no contribution to $\omega$. The different contribution in the equation are then 
\bea
 \star d \omega_1 &=& d \Big( \frac{1}{2 r \Sigma} \Big) - \frac{1}{r} d \frac{1}{\Sigma} \,, \qquad
 \star d \omega_2 = d \Big( \frac{\cos \theta}{R r \Sigma} \Big) - \frac{1}{r^2} d \frac{1}{\Sigma} \,, \\ \nonumber
 && \qquad \qquad \qquad  \star d \omega_3 = d  \Big( -\frac{\cos \theta_\Sigma}{2 R r^2} \Big) - \frac{1}{r \Sigma} d \frac{1}{r} \,. 
\eea
Again by a brute force resolution, the solutions are
\bea
 \omega_1 = \frac{ R \cos \theta - r }{ 2 R \Sigma } d\phi \,, \quad \omega_2 = -\frac{\sin^2 \theta }{ R \Sigma } d\phi \,, \quad
 \omega_3 = \frac{ \sin^2 \theta }{ 2 R \Sigma } d\phi \,.
\eea
This leads to the full solution for $\omega$
\bea
 \omega &=& \Big[ \, \kappa + \Big( m - \frac{c^{ijk}}{2} l_i l_j d_k \Big) \cos\theta + \Big( \ti{m} - \frac{c^{ijk}}{2} l_i l_j \ti{d}_k \Big) \cos\theta_\Sigma -\alpha \frac{\sin^2\theta}{r} - \ti{\alpha} \frac{\sin^2\theta_\Sigma}{\Sigma} \nonumber  \\
 && + c^{ijk} \Big( l_i( \ti{Q}_j d_k - Q_j \ti{d}_k ) \frac{r - R \cos\theta}{2 R \Sigma} - Q_i Q_j \ti{d}_k \frac{\sin^2 \theta}{2 R \Sigma} \\ \nonumber 
 && + \ti{Q}_i \ti{Q}_j d_k \frac{r \sin^2\theta}{2 R \Sigma^2} + Q_i \ti{Q}_j d_k \frac{\sin^2 \theta}{2 R \Sigma} - \ti{Q}_i Q_j \ti{d}_k \frac{r \sin^2\theta}{2 R \Sigma^2} \, \Big) \, \Big] d\phi \,.
\eea
The vectors $v_\Lambda$ are then easy to find. Indeed, from \eqref{EMvectorssm} $v_0$ is just equal to $\omega$, and the $v_i$'s are the harmonic duals of the $L_i$'s functions. Then we have
\bea
 v_0 &=& \omega \,, \\ \nonumber
 v_i &=& Q_i \cos\theta + \ti{Q}_i \cos\theta_\Sigma \,, \ i=1,2,3 \,. 
\eea
As for $V$, finding the solution for the $w^\Lambda$ does not involve any new terms, and the solutions are given by
\bea
 w^0 &=& \Big[ -(Q_6 + c^{ijk} k_i k_j Q_k) \cos\theta - (\ti{Q}_6 + c^{ijk} k_i k_j \ti{Q}_k) \cos\theta_\Sigma + \beta \frac{\sin^2\theta}{r} + \ti{\beta} \frac{\sin^2\theta_\Sigma}{\Sigma} \nonumber  \\
 && + 2 c^{ijk} \Big( k_i( \ti{d}_j Q_k - d_j \ti{Q}_k ) \frac{r - R \cos\theta}{2 R \Sigma} + d_i d_j \ti{Q}_k \frac{\sin^2 \theta}{R \Sigma} \\ \nonumber 
 && - \ti{d}_i \ti{d}_j Q_k \frac{r \sin^2\theta}{R \Sigma^2} - d_i \ti{d}_j Q_k \frac{\sin^2 \theta}{2 R \Sigma} + \ti{d}_i d_j \ti{Q}_k \frac{r \sin^2\theta}{2 R \Sigma^2} \, \Big) \, \Big] d\phi \,, \\ \nonumber
 w^i \! &=& \omega \! - \! c^{ijk} \left( \! (k_j Q_k - l_j d_k) \cos\theta \! + \! (k_j \ti{Q}_k - l_j \ti{d}_k) \cos\theta_\Sigma \! + \! (d_j \ti{Q}_k - \ti{d}_j Q_k) \frac{r - R\cos\theta}{R \Sigma} \right) \! d\phi \,.
\eea
%

\subsection{Regularity and physical properties}

The solution written in the previous subsection solves the equations of motion, but does not {\it a priori} satisfies the regularity conditions, necessary for the solution to be physical. We investigate this question in this subsection. One notes that the regularity conditions are exactly the same in four and five dimensions. However, the solution we construct being asymptotically Minkowski in four dimensions, the physical charges are the four-dimensional ones. 

\paragraph {Dirac--Misner strings.}

We will first implement the conditions for the absence of Dirac--Misner strings singularities. This requires $\omega$ to vanish for $\sin\theta=0$. Because of the presence of the two centres, this imposes three conditions, 
\bea
 \omega_{\theta=\pi} = 0 \,, \quad \omega_{\theta=0, \, r<R } = 0 \,, \quad \omega_{\theta=0, \, r>R} = 0 \,. 
\eea
They can be solved by affecting the values of $\kappa$, $m$ and $\ti{m}$ to 
\bea
 \kappa = \frac{c^{ijk}}{2R}  l_i \left( Q_j \ti{d}_k - \ti{Q}_j d_k \right) \,, \quad m = -\kappa + \frac{c^{ijk}}{2} l_i l_j d_k \,, \quad
 \ti{m} = \kappa + \frac{c^{ijk}}{2} l_i l_j \ti{d}_k \,.
\eea
Solving for $\kappa$, one can rewrite it as
\bea
 m + \ti{m} &=& \frac{c^{ijk}}{2} l_i l_j (d_k + \ti{d}_k) \,, \\ 
 -m + \frac{c^{ijk}}{2} l_i l_j d_k &=& \frac{c^{ijk}}{2R}  l_i \left( Q_j \ti{d}_k - \ti{Q}_j d_k \right) \,.
\eea
The second equation is similar to the non-BPS version of the integrability equation, or bubble equation \cite{Denef,Bates,Bena:2005va,Berglund:2005vb,Saxena:2005uk}. However, it should not be understood as such. Indeed, using (\ref{EMvectorssm}) one sees that this condition is equivalent as to the vanishing of the D0 charge of the two black holes. Nevertheless, we will show in the following that the distance between the two centres is determined in function of the electromagnetic charges of the two black hole constituents and the asymptotic value of the scalar fields. The property that this constraint does not directly follow from the absence of Dirac--Misner string singularities is peculiar to the simple duality frame we are considering, and we expect it would be fixed in this way as in the BPS case if we were considering the most general solution, function of the three phases $\alpha_i$ defining the duality frame.

\paragraph {Horizon regularity.}

We then have to make sure that each of the centre is a regular black hole. We perform this analysis at $r\to 0$, the result at $\Sigma\to 0$ being exactly equivalent. The horizon area is given by
\be
 A = \int d\theta d\phi \sqrt{r^2 (\rm{e}^{-4U} r^2 \sin^2\theta - \omega^2)} 
\ee
and it is therefore necessary for this quantity to be finite as $r \to 0$. One has
\bea
 \omega \stackrel{r\to 0}{=} - \alpha \frac{\sin^2\theta}{r} + \mathcal{O}( r^0) \,, 
\eea
which implies that the contribution from $\omega^2$ is always finite and well-behaved. On the contrary, as $r\to 0$, e$^{-4U}$ diverges in general as $1/r^6$ :
\bea
 \rm{e}^{-4U} \stackrel{r\to 0}{=} \frac{1}{r^6}\left( \frac{1}{6}Q_1 Q_2 Q_3 (c^{ijk}Q_i d_j d_k) - \frac{1}{36} (c^{ijk} Q_i Q_j d_k)^2 \right) + \mathcal{O}\big( r^{-5} \big) \,.
\eea
This term vanishes if 
\be
\frac{d_1}{Q_1} = \frac{d_2}{Q_2} = \frac{d_3}{Q_3} = \gamma \,,
\ee
for some arbitrary constant $\gamma$. With this condition, e$^{-4U}$ still diverges as $1/r^5$ :
\be
 \rm{e}^{-4U} \stackrel{r\to 0}{=} \frac{Q_1 Q_2 Q_3 (\beta - 2 \gamma \alpha ) \cos\theta}{r^5} +{\cal O}\big(r^{-4} \big)\,,
\ee
which imposes us to take
\be
 \beta = 2\gamma \alpha \,.
\ee
This conditions are now enough to ensure that the black hole located at $r=0$ has a finite entropy, and one shows that they are necessary. The argumentation is exactly the same for the centre at $\Sigma=0$, and therefore we also imposes
\bea
 \frac{\ti{d}_1}{\ti{Q}_1} &=& \frac{\ti{d}_2}{\ti{Q}_2} = \frac{\ti{d}_3}{\ti{Q}_3} = \ti{\gamma} \,, \\ \nonumber 
 \ti{\beta} &=& 2\ti{\gamma} \ti{\alpha} \,.
\eea
With these conditions, the entropy of the black hole located at $r=0$ is finally given by 
\bea \label{entropy}
 S_{r=0} &=& \pi \sqrt{ - I_{r=0} - \alpha^2} \,, \quad \rm{with} \\
-  I_{r=0} &\equiv& Q_1 Q_2 Q_3 \left( \! Q_6 - 2 \gamma m + \frac{c^{ijk}}{2} Q_i \left( \! (k_j + \gamma l_j)(k_k + \gamma l_k) -\frac{(\gamma - \ti{\gamma})^2}{R^2} \ti{Q}_j \ti{Q}_k \!\right) \!\right) \nonumber
\eea

We finally need to make sure that there are no closed time-like curves in the solution. This is done by imposing e$^{-4U}$ and $\rm{e}^{-4U} r^2 \sin^2\theta - \omega^2$ to be positive everywhere. Given the complicated explicit expressions for this quantities, we have not been able to find analytic conditions for their positivity. Nonetheless, a careful numerical analysis in different cases shows that regularity near each of the centres seems to be sufficient to have the global positivity everywhere. 

\paragraph{Mass, Charges, and angular momentum.}

One can now give the charges associated to our solution.
The total electromagnetic charges are given by 
\bea \label{defcharges}
 p^\Lambda &=& - \frac{1}{4\sqrt{2} \pi}\int_{S^2_\infty} d w^\Lambda = \frac{1}{2\sqrt{2}}  \bigl( w^\Lambda_\varphi(\theta_\Sigma= 0) - w^\Lambda_\varphi(\theta= \pi) \bigr) 
 \,, \CR
  q_\Lambda &=& - \frac{1}{4\sqrt{2} \pi}\int_{S^2_\infty} d v_\Lambda = \frac{1}{2\sqrt{2}}  \bigl( v_{\Lambda \varphi}(\theta_\Sigma= 0) - v_{\Lambda \varphi}(\theta= \pi) \bigr) 
 \,.
\eea
This gives
\bea
 p^0 &=& -\frac{1}{\sqrt{2}} \left( (Q_6 + \ti{Q}_6) + c^{ijk} k_i k_j (Q_k + \ti{Q}_k) \right) \,,  \quad \quad \quad \quad  q_0 = 0 \,, \\ \nonumber
 p^i &=& \frac{c^{ijk}}{\sqrt{2}} \left( l_j (d_k + \ti{d}_k) - k_j (Q_k + \ti{Q}_k) \right) \,,   \quad   q_i = \frac{1}{\sqrt{2}} ( Q_i + \ti{Q}_i ) \,.
\eea
In terms of branes, we recall that $p^0$ corresponds to the \Db6 charge, the $p^i$'s to the \Db{4}'s, the $q_i$'s to the D2's and $q_0$ to the D0. 

It will also be interesting to look at the individual charges of each centre. They are easily computed as in \eqref{defcharges}, by picking a cycle surrounding one of the centres.\footnote{\eg $ p^\Lambda_\pA =  \frac{1}{2\sqrt{2}}  \bigl( w^\Lambda_\varphi(\theta=0,\theta_\Sigma= \pi) - w^\Lambda_\varphi(\theta= \pi) \bigr) $ and $ p^\Lambda_\pB =  \frac{1}{2\sqrt{2}}  \bigl( w^\Lambda_\varphi(\theta_\Sigma= 0) - w^\Lambda_\varphi(\theta=0,\theta_\Sigma= \pi) \bigr) $.}  The electromagnetic charges of the first and the second centre, denoted respectively as $(p^\Lambda_\pA,q_{\pA\, \Lambda})$ and $(p^\Lambda_\pB,q_{\pB\, \Lambda})$, are given by
\bea \label{LocalCharges}
 q_{\pA\, i} &=& \frac{1}{\sqrt{2}} Q_i \; , \qquad p_\pA^i = - \frac{1}{\sqrt{2}} c^{ijk} \Bigl( k_j - \gamma l_j - \frac{\gamma - \tilde \gamma}{R} \tilde Q_j \Bigr) Q_l \CR
p^0_\pA &=& - \frac{1}{\sqrt{2}} Q_6 - \frac{1}{\sqrt{2}} c^{ijk} k_i \Bigl( k_j -  \frac{\gamma - \tilde \gamma}{R} \tilde Q_j \Bigr) Q_k \CR
q_{\pB\, i} &=& \frac{1}{\sqrt{2}} \tilde Q_i \; , \qquad p_\deux^i = - \frac{1}{\sqrt{2}} c^{ijk} \Bigl( k_j -\tilde  \gamma l_j - \frac{\tilde \gamma -  \gamma}{R}  Q_j \Bigr) \tilde Q_l \CR
p^0_\pB &=& - \frac{1}{\sqrt{2}} \tilde Q_6 - \frac{1}{\sqrt{2}} c^{ijk} k_i \Bigl( k_j -  \frac{\tilde \gamma - \gamma}{R}  Q_j \Bigr) \tilde Q_k \,.
\eea 
Having computed these charges, one can easily check that, as expected, the entropy of each centre \eqref{entropy} takes the conventional single-centre expression in terms of local charges and the intrinsic angular momenta, \ie 
\be
 S_{r=0} = \pi \sqrt{-I_4(q_{\pA\, \Lambda},p^\Lambda_\pA) - \alpha^2} \,, 
\ee
where $I_4(q,p)$ is the quartic invariant defined in \eqref{quarticinvariant}. The equivalent formula holds for the entropy of the second centre. 

The expression of the local charges also allows us to obtain the constraint giving the distance $R$ between the two centres as a function of them. Its explicit value is a rather complicated function of the charges and the asymptotic moduli. Defining the quantity 
\be
 \vartheta_i \equiv \frac{ q_{\pB\, i} p^i_\pB }{q_{\pB\, i+1} q_{\pB\, i+2}} - \frac{p^{i+1}_\pB}{q_{\pB\, i+2}} - \frac{p^{i+2}_\pB}{q_{\pB\, i+1}} - \frac{ q_{\pA\, i} p^i_\pA }{q_{\pA\, i+1} q_{\pA\, i+2}} + \frac{p^{i+1}_\pA}{q_{\pA\, i+2}} +\frac{p^{i+2}_\pA}{q_{\pA\, i+1}}\; , 
\ee
where no sum over $i$ is involved, one can show that 
\be
 2( \gamma - \tilde \gamma) \Bigl( l_i + \sqrt{2} \frac{ q_i }{R} \Bigr) = \vartheta_i \; , 
\ee
for all $i$. This gives that 
\be \label{RfucC} 
 R =  \sqrt{2}\,\frac{ q_{i+1} \vartheta_{i+2} - q_{i+2} \vartheta_{i+1}}{l_{i+2} \vartheta_{i+1} - \vartheta_{i+2} l_{i+1} }   
\ee
whatever the choice of $i$ is, and with no sum on $i$. Note that this equation implies therefore two extra conditions on the charges of the constituent black holes and the asymptotic moduli. Together with the condition that $q_0 = 0 $, the system requires three constraints which correspond to the fix value of the phases $\alpha_i$ defining $h = 2 \sum_i H_i$ in this duality frame. It is therefore difficult to interpret the constraint (\ref{RfucC}) as coming from the absence of time-like closed curves in this duality frame, and the definition of the correct  integrability equation, or bubble equation \cite{Denef,Bates,Bena:2005va,Berglund:2005vb,Saxena:2005uk},  would require to consider the general solution with three arbitrary phases. 

\medskip

We now want to compute the mass and angular momentum of the solution. In order to do this, it is convenient to solve for $h$ in function of the other constants by imposing asymptotic flatness, \ie  $U\rightarrow 0$
\be
  h  = \frac{1 + m_0^{\; 2}}{l_1 l_2 l_3} -  \sum_{i} l_i k_{i+1} k_{i+2}\; . 
\ee
For convenience we also define the quantities $f_i$ 
\be
 f_i \equiv k_i l_{i+1} l_{i+2}  - m_0 \; , 
\ee
such that the asymptotic of the scalar fields reduces to  \footnote{One has reciprocally $f_i = e^{2\phi_i} a_i$ and $l_i^{\; 2} = \frac{ \prod_{j\ne i} ( e^{-2\phi_j} + e^{2\phi_j} a_j^{\, 2}) }{ e^{-2\phi_i} + e^{2\phi_i} a_i^{\, 2} }$.}
\be
 t^i_\asym = l_{i+1} l_{i+2} \frac{f_i + i }{1 + f_i^{\; 2}} \; . 
\ee
In particular, the axions vanish for $f_i=0$. One computes the ADM mass 
\be \label{massformula}
 M = \frac{1}{2\sqrt{2}} \Bigl( -  l_1 l_2 l_3 p^0 + \sum_i f_i l_i p^i + \sum_i \frac{1 + f_{i+1} f_{i+2}}{l_i} q_i \Bigr) \; . 
\ee
This formula reproduce the ADM mass as given in (\ref{nonBPSADM}) for the value of the phase $\alpha_i$ 
\be \label{angleCond}
 e^{-i \alpha_i} = \frac{ ( i - f_{i} ) ( i + f_{i+1} ) ( i + f_{i+2} ) }{\prod_j \sqrt{ 1 + f_j^{\; 2}}} \; . 
\ee 
One important thing to note is that the property that the phases only depend on the asymptotic moduli $t^i_\asym$ is due to the simplicity of the specific duality frame we chose; and that in general these phases depend non-trivially on the electromagnetic charges, as they do for single-centre solutions in particular. Although the mass appears to be linear in the electromagnetic charges in this particularly simple duality frame, this is a specific property of the latter and the mass formula would be highly non-linear for more general charges and asymptotic moduli. 

In fact it is important for the stability that this mass formula does not apply for the single-centre solutions with electromagnetic charges $q_{\pA\, \Lambda}, p_\pA^\Lambda$ and $q_{\pB\, \Lambda}, p_\pB^\Lambda$, because one would then conclude that the mass of the composite solution is equal to the sum of the masses of the constituent black holes, and thus that the solution is only marginally stable. Note indeed that one cannot cancel the charges of one centre 
 in (\ref{LocalCharges}) without modifying the electromagnetic charges of the other, such that single-centre solutions with the charges (\ref{LocalCharges}) are not themselves solutions of the system in this particular duality frame. One computes instead that the mass formula reproduces the value of the non-BPS fake superpotential if and only if 
\be p^i = - \frac{1}{l_i} \Scal{ ( f_{i+1} + s )  \frac{q_{i+2}}{l_{i+2}} + ( f_{i+2} + s )  \frac{q_{i+1}}{l_{i+1} }}\; ,  \ee
which is the case when $R\rightarrow \infty$ for $s = m_0 - \gamma l_1 l_2 l_3$. For more general charges, and in particular when $R$ is finite and the solution is expected to be stable, the mass formula (\ref{massformula}) corresponds to the `auxiliary fields' dependent generalised fake superpotential \cite{Ceresole:2009vp} for which the auxiliary scalars, say $\beta^s_\alpha$ for $\alpha=1,2$, are not extremum of the generalised fake superpotential, $ \frac{\partial W(\beta)}{\partial \beta_\alpha }\big|_{\beta = \beta^s}  \ne 0 $. The auxiliary scalars $\beta_\alpha$ parametrize the three phases $\alpha_i$ satisfying that (\ref{nonBPSADM}) is real and positive, and $\beta^s_\alpha$ are determined by the conditions (\ref{angleCond}), and therefore only depend on the asymptotic value of the scalar fields, and not the electromagnetic charges. The true `fake superpotential' (which defines the ADM mass of a single-centre non-BPS black hole) is instead defined as $W(\beta^*)$, for which $\frac{\partial W(\beta)}{\partial \beta_\alpha }\big|_{\beta = \beta^*} =  0 $. As advocated in \cite{Ceresole:2009vp}, $W(\beta^*)$ is bounded from below as a function of the moduli, in which case it coincides with the expression determined in \cite{BossardW}, whereas it develops unstable directions for other values of the auxiliary scalars $\beta_\alpha$. It must therefore exist configurations such that $W(\beta^s) < W(\beta^*)$, but this is not true in general. In the present situation, the solution will be stable if 
\begin{multline}  M = W(\beta^s,q_\pA+q_\pB,p_\pA+p_\pB) = W(\beta^s,q_\pA,p_\pA) + W(\beta^s,q_\pB,p_\pB) \\ <  \; W(\beta^*_\pA,q_\pA,p_\pA)+ W(\beta^*_\pB,q_\pB,p_\pB) = M_\pA + M_\pB  \; . \end{multline}
This is trivially satisfied if both $\beta^*_\pA$ defines a maximum of $W(\beta,q_\pA,p_\pA)$ and $\beta^*_\pB$ a maximum of $W(\beta,q_\pB,p_\pB)$, but this inequality does not seem to be satisfied for arbitrary charges and moduli, as opposed to the BPS case for which the triangle inequality $|Z(q_\pA+q_\pB,p_\pA+p_\pB)| \le |Z(q_\pA,p_\pA)|+|Z(q_\pB,p_\pB)|$ holds in general.  

The angular momenta of the solution is given by
\bea
  J &=& - (\alpha + \ti{\alpha}) + \frac{\gamma - \ti{\gamma}}{2} c^{ijk} \Big( l_i + \frac{Q_i + \ti{Q}_i}{R} \Big) Q_j \ti{Q}_k \CR
 &=&  - (\alpha + \ti{\alpha} ) + p_\pA^\Lambda q_{\pB\, \Lambda} -  q_{\pA\, \Lambda}\,  p_\pB^\Lambda  \; .  
\eea
where in the second line we used the definition of the individual charges of each centre. It is very interesting to note that, when written in terms of the physical charges, the angular momentum takes the very natural form of the sum of the intrinsic angular momenta plus the contribution coming from the electromagnetic interaction between the centres.

\bigskip

\bigskip
\section*{Acknowledgments}

We would like to thank Iosif Bena, Stefano Giusto, Josef Lindman H\"ornlund, Boris Pioline, Amitabh Virmani and Nick Warner for fruitful discussions. The work of G.B. was supported by the ITN programme PITN-GA-2009-237920, the ERC Advanced Grant 226371, the IFCPAR CEFIPRA programme 4104-2 and the ANR programme blanc NT09-573739.

\smallskip
%

%
%

\appendix 

\section{Conventions for $SO(4,4)$}
\renewcommand{\theequation}{A.\arabic{equation}}
\setcounter{equation}{0}

In this appendix we give the explicit form of some of the conventions presented in section~\ref{sec3}. 

First of all, our representation of $SO(4,4)$ is defined as the one preserving the matrix \eqref{defeta}, it is explicitly given by
\be 
\upeta =   
\left(\begin{array}{cccccccc} 
\hspace{2mm} 0 \hspace{2mm}& \hspace{2mm} 0 \hspace{2mm}  & \hspace{2mm} 0 \hspace{2mm} & \hspace{2mm} 1 \hspace{2mm} & \hspace{2mm} 0 \hspace{2mm} & \hspace{2mm} 0 \hspace{2mm} & \hspace{2mm} 0 \hspace{2mm} & \hspace{2mm} 0 \hspace{2mm} \\
\hspace{2mm} 0 \hspace{2mm}& \hspace{2mm} 0 \hspace{2mm}  & \hspace{1mm}  -1 \hspace{2mm} & \hspace{2mm} 0 \hspace{2mm} & \hspace{2mm} 0 \hspace{2mm} & \hspace{2mm} 0 \hspace{2mm} & \hspace{2mm} 0 \hspace{2mm} & \hspace{2mm} 0 \hspace{2mm} \\
\hspace{2mm} 0 \hspace{2mm}& \hspace{1mm} -1 \hspace{2mm}  & \hspace{2mm} 0 \hspace{2mm} & \hspace{2mm} 0 \hspace{2mm} & \hspace{2mm} 0 \hspace{2mm} & \hspace{2mm} 0 \hspace{2mm} & \hspace{2mm} 0 \hspace{2mm} & \hspace{2mm} 0 \hspace{2mm} \\
\hspace{2mm} 1 \hspace{2mm}& \hspace{2mm} 0 \hspace{2mm}  & \hspace{2mm} 0 \hspace{2mm} & \hspace{2mm} 0 \hspace{2mm} & \hspace{2mm} 0 \hspace{2mm} & \hspace{2mm} 0 \hspace{2mm} & \hspace{2mm} 0 \hspace{2mm} & \hspace{2mm} 0 \hspace{2mm} \\
\hspace{2mm} 0 \hspace{2mm}& \hspace{2mm} 0 \hspace{2mm}  & \hspace{2mm} 0 \hspace{2mm} & \hspace{2mm} 0 \hspace{2mm} & \hspace{2mm} 0 \hspace{2mm} & \hspace{2mm} 0 \hspace{2mm} & \hspace{2mm} 0 \hspace{2mm} & \hspace{2mm} 1 \hspace{2mm} \\
\hspace{2mm} 0 \hspace{2mm}& \hspace{2mm} 0 \hspace{2mm}  & \hspace{2mm} 0 \hspace{2mm} & \hspace{2mm} 0 \hspace{2mm} & \hspace{2mm} 0 \hspace{2mm} & \hspace{2mm} 0 \hspace{2mm} & \hspace{1mm} -1 \hspace{2mm} & \hspace{2mm} 0 \hspace{2mm} \\
\hspace{2mm} 0 \hspace{2mm}& \hspace{2mm} 0 \hspace{2mm}  & \hspace{2mm} 0 \hspace{2mm} & \hspace{2mm} 0 \hspace{2mm} & \hspace{2mm} 0 \hspace{2mm} & \hspace{1mm} -1 \hspace{2mm} & \hspace{2mm} 0 \hspace{2mm} & \hspace{2mm} 0 \hspace{2mm} \\
\hspace{2mm} 0 \hspace{2mm}& \hspace{2mm} 0 \hspace{2mm}  & \hspace{2mm} 0 \hspace{2mm} & \hspace{2mm} 0 \hspace{2mm} & \hspace{2mm} 1 \hspace{2mm} & \hspace{2mm} 0 \hspace{2mm} & \hspace{2mm} 0 \hspace{2mm} & \hspace{2mm} 0 \hspace{2mm} 
\end{array} \right) \,.
\ee 
The algebra elements appearing in \eqref{defV}
\bea 
   \V &=&  \exp\bigl[ \zeta^{A_1B_2C_3} {\bf E}_{A_1B_2C_3} + \sigma {\bf E} \bigr] \exp[ U {\bf H}] 
\left(\begin{array}{ccc} 
\hspace{2mm} v_{a_1}{}^{A^\prime_1}  \hspace{2mm}& \hspace{2mm} 0  \hspace{2mm}  & \hspace{2mm}0 \hspace{2mm}  \\
\hspace{2mm}0  \hspace{2mm}& \hspace{2mm} v_{a_1}{}^{A_1^\prime}  \hspace{2mm}  & \hspace{2mm} 0  \hspace{2mm}  \\
\hspace{2mm}  0    \hspace{2mm}& \hspace{2mm}  0 \hspace{2mm}  & \hspace{2mm} (v^{-1})_{B_2^\prime}{}^{b_2} (v^{-1})_{C_3^\prime}{}^{c_3} \hspace{2mm}
 \end{array} \right)  \begin{array}{c} \\ \\ \vspace{4mm} \end{array}
\eea
are 
\be
 \zeta^{A_1B_2C_3} {\bf E}_{A_1B_2C_3} + \sigma {\bf E} = 
\left(
\begin{array}{cccccccc}
 0 \ \ & 0 \ & \sigma  & 0 & \psi_0 & -\psi_3 & -\psi_2 & -\chi^1 \\
 0 \ \ & 0 \ & 0 & \sigma  & -\psi_1 & -\chi^2 & -\chi^3 & -\chi^0 \\
 0 \ \ & 0 \ & 0 & 0 & 0 & 0 & 0 & 0 \\
 0 \ \ & 0 \ & 0 & 0 & 0 & 0 & 0 & 0 \\
 0 \ \ & 0 \ & -\chi^0 & \chi^1 & 0 & 0 & 0 & 0 \\
 0 \ \ & 0 \ & \chi^3 & -\psi_2 & 0 & 0 & 0 & 0 \\
 0 \ \ & 0 \ & \chi^2 & -\psi_3 & 0 & 0 & 0 & 0 \\
 0 \ \ & 0 \ & -\psi_1 & -\psi_0 & 0 & 0 & 0 & 0
\end{array}
\right) \,,
\ee
\be
 U {\bf H} = 
\left(
\begin{array}{cccccccc}
 U \ & 0 & 0 & 0 \ & 0 \ \ & 0 \ \ & 0 \ \ & 0 \\
 0 \ & U & 0 & 0 \ & 0 \ \ & 0 \ \ & 0 \ \ & 0 \\
 0 \ & 0 & -U & 0 \ & 0 \ \ & 0 \ \ & 0 \ \ & 0 \\
 0 \ & 0 & 0 & -U \ & 0 \ \ & 0 \ \ & 0 \ \ & 0 \\
 0 \ & 0 & 0 & 0 \ & 0 \ \ & 0 \ \ & 0 \ \ & 0 \\
 0 \ & 0 & 0 & 0 \ & 0 \ \ & 0 \ \ & 0 \ \ & 0 \\
 0 \ & 0 & 0 & 0 \ & 0 \ \ & 0 \ \ & 0 \ \ & 0 \\
 0 \ & 0 & 0 & 0 \ & 0 \ \ & 0 \ \ & 0 \ \ & 0
\end{array}
\right) \; , 
\ee
and
\begin{multline}
 \left(\begin{array}{ccc} 
\hspace{2mm} v_{a_1}{}^{A^\prime_1}  \hspace{2mm}& \hspace{2mm} 0  \hspace{2mm}  & \hspace{2mm}0 \hspace{2mm}  \\
\hspace{2mm}0  \hspace{2mm}& \hspace{2mm} v_{a_1}{}^{A_1^\prime}  \hspace{2mm}  & \hspace{2mm} 0  \hspace{2mm}  \\
\hspace{2mm}  0    \hspace{2mm}& \hspace{2mm}  0 \hspace{2mm}  & \hspace{2mm} (v^{-1})_{B_2^\prime}{}^{b_2} (v^{-1})_{C_3^\prime}{}^{c_3} \hspace{2mm}
 \end{array} \right)  \begin{array}{c} \\ \\ \vspace{4mm} \end{array}  =
 \\ \nonumber \vspace{2mm}
  \left(
\begin{array}{cccccccc}
 e^{-\phi_1} \ \  & a_1 e^{\phi_1} & 0 \ \ & 0 & 0 & 0 & 0 & 0 \\
 0 \ \ & e^{\phi_1} & 0 \ \ & 0 & 0 & 0 & 0 & 0 \\
 0 \ \ & 0 & e^{-\phi_1} \ \ & a_1 e^{\phi_1} & 0 & 0 & 0 & 0 \\
 0 \ \ & 0 & 0 \ \ & e^{\phi_1} & 0 & 0 & 0 & 0 \\
 0 \ \ & 0 & 0 \ \ & 0 & e^{\phi_2+\phi_3} & 0 & 0 & 0 \\
 0 \ \ & 0 & 0 \ \ & 0 & -a_3 e^{\phi_2+\phi_3} & e^{\phi_2-\phi_3} & 0 & 0 \\
 0 \ \ & 0 & 0 \ \ & 0 & -a_2 e^{\phi_2+\phi_3} & 0 & e^{-\phi_2+\phi_3} & 0 \\
 0 \ \ & 0 & 0 \ \ & 0 & a_2 a_3 e^{\phi_2+\phi_3} & -a_2 e^{\phi_2-\phi_3} & -a_3 e^{-\phi_2+\phi_3} & e^{-\phi_2-\phi_3}
\end{array}
\right) \,.
\end{multline}

\medskip

The solution of (\ref{Heigen}) in terms of the matrix (\ref{CentralCharge}) is 
 \be \label{D0D2D4D6}
\begin{split} 
 D0 \ : \   &(1,1,1,1) \\
 D2 \ : \ &  (1,-1,1,1) \\
 &  (1,1,-1,1) \\
 &  (1,1,1,-1) \\
 D4 \ : \ &  (1,1,-1,-1)\\
 &  (1,-1,1,-1) \\
 &  (1,-1,-1,1) \\
 D6 \ : \ &  (1,-1,-1,-1)\\
 \overline{D0} \ : \   &(-1,-1,-1,-1) \\
 \overline{D2} \ : \ &  (-1,1,-1,-1) \\
 &  (-1,-1,1,-1) \\
 &  (-1,-1,-1,1) \\
 \overline{D4} \ : \ &  (-1,-1,1,1)\\
 &  (-1,1,-1,1) \\
 &  (-1,1,1,-1) \\
 \overline{D6} \ : \ &  (-1,1,1,1)
 \end{split}
\hspace{10mm}
\begin{split}
 -N &= - \Xi_i =  Q_i = - P^0 \\
 M &= - \Sigma_1 = \Sigma_2 = \Sigma_3 =- \stfrac{1}{2} P^1 \\
 M &=  \Sigma_1 = -\Sigma_2 = \Sigma_3 = -\stfrac{1}{2} P^2 \\
 M &=  \Sigma_1 = \Sigma_2 = -\Sigma_3 = -\stfrac{1}{2} P^3 \\
 -N &= - \Xi_1 = \Xi_2 = \Xi_3 = -Q_1 =  Q_2 = Q_3 = P^0  \\
 -N &=  \Xi_1 = -\Xi_2 = \Xi_3 = Q_1 =  -Q_2 =  Q_3 = P^0  \\
 -N &=  \Xi_1 = \Xi_2 = -\Xi_3 = Q_1 =  Q_2 =  -Q_3 = P^0  \\
 M &= - \Sigma_i = \stfrac{1}{2} Q_0 \\
 -N &= - \Xi_i = -Q_i =  P^0 \\
 M &= - \Sigma_1 = \Sigma_2 = \Sigma_3 =   \stfrac{1}{2} P^1 \\
 M &=  \Sigma_1 = -\Sigma_2 = \Sigma_3 = \stfrac{1}{2} P^2 \\
 M &=  \Sigma_1 = \Sigma_2 = -\Sigma_3 =  \stfrac{1}{2} P^3 \\
 -N &= - \Xi_1 = \Xi_2 = \Xi_3 =  Q_1 = - Q_2 = - Q_3 = - P^0  \\
 -N &=  \Xi_1 = -\Xi_2 = \Xi_3 = -Q_1 = Q_2 =  -Q_3 = -P^0  \\
 -N &=  \Xi_1 = \Xi_2 = -\Xi_3 = -Q_1 =  -Q_2 =  Q_3 = -P^0  \\
 M &= - \Sigma_i = - \stfrac{1}{2} Q_0 
\end{split}
\ee
These definitions deserve some explanations. The four-uplet of numbers give the weight of each generator with respect to the four generators $H_\Lambda$. In particular, given these numbers, one can just read the weight of the generators with respect to $h=b^\Lambda H_\Lambda$. In particular, for the BPS orbit, $b^\Lambda=(2,0,0,0)$ and thus the weight is the first number of the four-uplet. In the duality frame that we have chosen in this paper, one can then show that each of the generator of the first part of the list, that all have weight one, couple respectively to D0, D2, D4, and D6-brane charges. The $\overline{\rm{D}}$-brane generators are then naturally the dual one, with opposite weights. From this weights, it is then straightforward to see that the almost-BPS orbit ($b^\Lambda=(4,2,2,2)$) corresponds to \Db{6}-D4-D2-D0 solutions and the composite non-BPS ($b^\Lambda=(0,2,2,2)$) to \Db{6}-\Db{4}-D2-D0 ones. 

\medskip

Finally, we give here the explicit form of the $\KK$-rotation matrix, that takes us from one duality frame to another. an element $R\in \KK$ acts on $\V\V^\ddagger$ as
\be
 \V \V^\ddagger \longrightarrow R \, \V \V^\ddagger \, R^T 
\ee
where $R$ is defined as
\be
 R = \left( \begin{array}{cc}
 R_4(\alpha_0' , \alpha_1') & 0 \\ 
 0 & R_4(\alpha_2' , \alpha_3') 
 \end{array} \right)
\ee
with
\be
 R_4(\alpha_i , \alpha_j) = \left( \begin{array}{cccc}
 \ \cos \alpha_i \cos \alpha_j \; & \ \cos \alpha_i \sin \alpha_j \; & \ \sin \alpha_i \cos \alpha_j \; & \ \sin \alpha_i \sin \alpha_j \\
 -\cos \alpha_i \sin \alpha_j \; & \ \cos \alpha_i \cos \alpha_j \; & -\sin \alpha_i \sin \alpha_j \; & \ \sin \alpha_i \cos \alpha_j \\
 -\sin \alpha_i \cos \alpha_j \; & - \sin \alpha_i \sin \alpha_j \; & \ \cos \alpha_i \cos \alpha_j \; & \ \cos \alpha_i \sin \alpha_j \\ 
  \ \sin \alpha_i \sin \alpha_j \; & - \sin \alpha_i \cos \alpha_j \; & - \cos \alpha_i \sin \alpha_j \; & \ \cos \alpha_i \cos \alpha_j 
 \end{array} \right) \,.
\ee
The angles $\alpha_\Lambda'$ appearing in the $R$ matrix are related to the one of \eqref{AMRotation} through

\be
 \alpha_0 = \alpha_0^\prime - \sum_i \alpha_i^\prime \; , \qquad \alpha_i = \alpha_0^\prime + \alpha_i^\prime \; . 
\ee



\end{document}